\documentclass[12pt,tightenlines,eqsecnum,floats,aps,amsmath,amssymb,nofootinbib,prd,showpacs]{revtex4}

\usepackage{setspace}
\usepackage{amsmath,amssymb,amsfonts,amsthm}
\usepackage{graphicx}
\usepackage{enumerate} % advanced enumerate environment
\usepackage{colordvi} % for color text

\def\be{\begin{equation}}
\def\ee{\end{equation}}
\def\ba{\begin{eqnarray}}
\def\ea{\end{eqnarray}}
\def\bi{\begin{itemize}}
\def\ei{\end{itemize}}
\def\bnum{\begin{enumerate}}
\def\enum{\end{enumerate}}

\def\V{\mathcal{V}}
\def\v{\nu}
\def\p{\phi}
\def\vp{\varphi}
\def\H{{\cal H}}
\def\Hkg{\H_{\rm kin}^{\rm grav}}
\def\Hk{\H_{\rm kin}}
\def\Hp{\H_{\rm phy}}
\def\Hkm{\H_{\rm kin}^{\rm matt}}

\def\wh{{}}
\def\b{\bar}

\def\f{\frac}

\def\w{\omega}

\def\lp{{\ell}_{\rm Pl}}
\def\l{\lambda}
\def\lo{\ell_o}
\def\bra{\langle}
\def\ket{\rangle}
\def\sint{\textstyle{\int}}

\def\R{\mathbb{R}}
\def\S{\mathcal{S}}
\def\Z{\mathbb{Z}}
\def\I{\mathcal{I}}

\def\dd{\textrm{d}}
\def\t{\tilde}
\def\b{\bar}
\usepackage{enumerate}

\usepackage{colordvi}
\newcounter{mnotecount}[section]

\newcommand{\comment}[1]{}
\begin{document}
\title{Casting Loop Quantum Cosmology in the Spin Foam Paradigm}

\author{Abhay Ashtekar}\email{ashtekar@gravity.psu.edu}
\author{Miguel Campiglia} \email{miguel@gravity.psu.edu}
\author{Adam Henderson}\email{henderson@gravity.psu.edu}
 \affiliation{Institute for Gravitation and the Cosmos \& Physics
 Department,
 Penn State, University Park, PA 16802-6300, U.S.A.}

\begin{abstract}

The goal of spin foam models is to provide a viable path integral
formulation of quantum gravity. Because of background independence,
their underlying framework has certain novel features that are not
shared by path integral formulations of familiar field theories in
Minkowski space. As a simple viability test, these features were
recently examined through the lens of loop quantum cosmology (LQC).
Results of that analysis, reported in a brief communication
\cite{ach1}, turned out to provide concrete arguments in support of
the spin foam paradigm. We now present detailed proofs of those
results. Since the quantum theory of LQC models is well understood,
this analysis also serves to shed new light on some long standing
issues in the spin foam and group field theory literature. In
particular, it suggests an intriguing possibility for addressing the
question of why the cosmological constant is positive and small.

\end{abstract}

% These were used for BKL: \pacs{04.20.Dw,04.60.Kz,04.60Pp,98.80Qc,04.20Fy}

\maketitle
\section{Introduction}
\label{s1}

Four different avenues to quantum gravity have been used to arrive
at  spin-foam models (SFMs). The fact that ideas from seemingly
unrelated directions converge to the same type of structures and
models has provided a strong impetus to the spin foam program over
the years \cite{cr-corfu}.

The first avenue is the Hamiltonian approach to loop quantum gravity
(LQG) \cite{alrev,crbook,ttbook}. By mimicking the procedure that
led Feynman \cite{rpf} to a sum over histories formulation of
quantum mechanics, Rovelli and Reisenberger \cite{rr} proposed a
space-time formulation of LQG. This work launched the spin-foam
program. The second route stems from the fact that the starting
point in canonical LQG is a rewriting of classical general
relativity that emphasizes connections over metrics
\cite{aa-newvar}. Therefore in the passage to quantum theory it is
natural to begin with the path integral formulation of appropriate
gauge theories. A particularly natural candidate is the topological
B-F theory \cite{gh} because in 3 space-time dimensions it is
equivalent to Einstein gravity, and in higher dimensions general
relativity can be regarded as a constrained BF theory \cite{jb-BF}.
The well-controlled path integral formulation of the BF theory
provided the second avenue and led to the SFM of Barret and Crane
\cite{bc}. The third route comes from the Ponzano-Regge model of
3-dimensional gravity \cite{pr} that inspired Regge calculus in
higher dimensions \cite{regge1,regge2,regge3}. Here one begins with
a simplicial decomposition of the space-time manifold, describes its
discrete Riemannian geometry using edge lengths and deficit angles
and constructs a path integral in terms of them. If one uses
holonomies and discrete areas of loop quantum gravity in place of
edge lengths, one is again led to a spin foam. These three routes
are inspired by various aspects of general relativity. The fourth
avenue starts from approaches to quantum gravity in which gravity is
to emerge from a more fundamental theory based on abstract
structures that, to begin with, have nothing to do with space-time
geometry. Examples are matrix models for 2-dimensional gravity and
their extension to 3-dimensions ---the Boulatov model \cite{bou}---
where the basic object is a field on a group manifold rather than a
matrix. The Boulatov  model was further generalized to a group field
theory (GFT) tailored to 4-dimensional gravity
\cite{crbook,gft1,gft2}. The perturbative expansion of this GFT
turned out be very closely related to the vertex expansions in SFMs.
Thus the SFMs lie at a junction where four apparently distinct paths
to quantum gravity meet. Through contributions of many researchers
it has now become an active research area (see, e.g.,
\cite{perezrev,jb-BF,crbook}).

Let us begin with the first path and examine SFMs from the
perspective of LQG. Recall that spin network states are used in LQG
to construct a convenient orthonormal basis in the kinematical
Hilbert space. A key challenge is to extract physical states from
them by imposing constraints. Formally this can be accomplished by
the group averaging procedure which also provides the physical inner
product between the resulting states \cite{dm,almmt}. From the LQG
perspective, the primary goal of SFMs is to construct a path
integral that leads to this physical Hilbert space.

Heuristically, the main idea behind this construction can be
summarized as follows. Consider a 4-manifold $M$ bounded by two
3-surfaces, $S_1$ and $S_2$, and a simplicial decomposition thereof.
One can think of $S_1$ as an `initial' surface and $S_2$ as a
`final' surface. One can fix a spin network on each of these
surfaces to specify an `initial' and a `final' state of the quantum
3-geometry. A quantum 4-geometry interpolating between the two can
be constructed by considering the dual triangulation of $M$ and
coloring its surfaces with half integers $j$ and edges with suitable
intertwiners. The idea is to obtain the physical inner product
between the two states by summing first over all the colorings for a
given triangulation, and then over triangulations keeping the
boundary states fixed. The second sum is often referred to as the
\emph{vertex expansion} because the $M$-th term in the series
corresponds to a dual triangulation with $M$ vertices. Since each
triangulation with a coloring specifies a quantum geometry, the sum
is regarded as a path integral over physically appropriate
4-geometries. In ordinary quantum mechanics and Minkowskian field
theories where we have a fixed background geometry, such a path
integral provides the (dynamically determined) transition amplitude
for the first state, specified at initial time, to evolve to the
second state at the final time. In the background independent
context of quantum gravity, one does not have access to a time
variable and dynamics is encoded in constraints. Therefore the
notion of a transition in a pre-specified time interval is not
a priori meaningful.
%
%\footnote{The notion \emph{is} meaningful if one can solve the
%quantum constraint by de-parametrization by introducing an
%appropriate relational time variable. We will return to this point
%below.}
%
Rather, the sum over histories now provides the physical inner
product between solutions to the quantum constraints, extracted from
the two spin network states.

Over the last two years there have been significant advances in
SFMs. While the structure of the path integral is well-motivated by
the interplay between general relativity and the BF theory, its
precise definition requires a key new ingredient ---the vertex
amplitude. The first proposal for the vertex amplitude was made over
ten years ago \cite{bc}. But it turned out to have important
limitations \cite{bcht,ar}. New proposals have now been put forward
\cite{epr,fk,eprl,ep} and, for the physically interesting regime of
the Barbero-Immirzi parameter, they agree. Furthermore, one can
regard these SFMs as providing an independent derivation of the
kinematics underlying LQG. The detailed agreement between LQG and
the new SFMs \cite{kkl1,kkl2} is a striking development. There are
also a number of results indicating that one does recover general
relativity in the appropriate limit \cite{limit1,limit2}. Finally,
the vertex amplitude is severely constrained by several general
requirements which the new proposals meet.

However, so far, the vertex amplitude has not been systematically
derived following procedures used in well-understood field theories,
or, starting from a well-understood Hamiltonian dynamics. Therefore,
although the convergence of ideas from several different directions
is impressive, a number of issues still remain. In particular, the
convergence is not quite as seamless as one would like; some rough
edges still remain because of unresolved tensions.

For example, the final vertex expansion is a discrete sum, in which
each term is itself a sum over colorings for a fixed triangulation.
A priori it is somewhat surprising that the final answer can be
written as a \emph{discrete} sum. Would one not have to take some
sort of a continuum limit at the end? One does this in the standard
Regge approach \cite{hamber} which, as we indicated above, is
closely related to SFMs. Another route to SFMs emphasizes and
exploits the close resemblance to gauge theories. In non-topological
gauge theories one also has to take a continuum limit. Why not in
SFMs? Is there perhaps a fundamental difference because, while the
standard path integral treatment of gauge theories is rooted in the
smooth Minkowskian geometry, SFMs must face the Planck scale
discreteness squarely?

A second potential tension stems from the fact that the construction of
the physical inner product mimics that of the transition amplitude
in Minkowskian quantum field theories. As noted above, in a
background independent theory, there is no a priori notion of time
evolution and dynamics is encoded in constraints. However, sometimes
it \emph{is} possible to `de-parameterize' the theory and solve the
Hamiltonian constraint by introducing an emergent or relational time
a la Leibnitz. What would then be the interpretation of the
spin-foam path integral? Would it yield both the physical inner
product \emph{and} the transition amplitude? Or, is there another
irreconcilable difference from the framework used Minkowskian field
theories?

There is a also a tension between SFMs and GFTs. Although fields in
GFTs live on an abstract manifold constructed from a Lie group, as
in familiar field theories the action has a free part and an
interaction term. The interaction term has a coupling constant,
$\lambda$, as coefficient. One can therefore carry out a Feynman
expansion and express the partition function, propagators, etc, as a
perturbation series in $\lambda$. If one sets $\lambda=1$, the
resulting series can be identified with the vertex expansion of
SFMs. But if one adopts the viewpoint that the GFT is fundamental
and regards gravity as an emergent phenomenon, one is led to allow
$\lambda$ to run under the renormalization group flow. What then is
the meaning of setting $\lambda=1$?  Or, do other values of
$\lambda$ have a role in SFMs that has simply remained unnoticed
thus far? Alternatively, one can put the burden on GFTs. They appear
to be efficient and useful calculational schemes. But if they are to
have a direct physical significance on their own, what then would the
\emph{gravitational} meaning of $\lambda$ be?

Such questions are conceptually and technically difficult. However,
they are important precisely because SFMs appear to lie at a
junction of several cross-roads and the recent advances bring out
their great potential. Loop quantum cosmology (LQC) provides a
physically interesting yet technically simple context to explore
such issues. In LQC the principles of LQG are applied to simple
cosmological models which have a high degree of symmetry. Thanks to
this symmetry, it has been possible to construct and analyze in
detail quantum theories in a number of cases
\cite{aps1,aps2,aps3,apsv,kv,acs,bp,ap,aps4,awe2,awe3,gowdy1,gowdy2,gowdy3}.
Furthermore, LQC shares many of the conceptual problems of LQG and
SFMs. Therefore it provides a fertile ground to test various ideas
and conjectures in the full theory. In the Hamiltonian context, LQC
has served this role successfully (for a recent review, see
\cite{aa-badhonef}). The goal of this paper is to first cast LQC in
the spin foam paradigm and then use the results to shed light on the
paradigm itself.

In LQC one can arrive at a sum over histories starting from a fully
controlled Hamiltonian theory. We will find that this sum bears out
the ideas and conjectures that drive the spin foam paradigm.
Specifically, we will show that: i) the physical inner product in
the timeless framework equals the transition amplitude in the theory
that is deparameterized using relational time; ii) this quantity
admits a vertex expansion a la SFMs in which the $M$-th term refers
just to $M$ volume transitions, without any reference to the time at
which the transition takes place; iii) the exact physical inner
product is obtained by summing over just the discrete geometries; no
`continuum limit' is involved; and, iv) the vertex expansion can be
interpreted as a perturbative expansion in the spirit of GFT, where,
moreover, the GFT coupling constant $\lambda$ is closely related to
the cosmological constant $\Lambda$. These results were reported in
the brief communication \cite{ach1}. Here we provide the detailed
arguments and proofs. Because the Hilbert space theory is fully
under control in this example, we will be able to avoid formal
manipulations and pin-point the one technical assumption that is
necessary to obtain the desired vertex expansion: one can interchange the group averaging integral and a convergent but infinite sum defining the gravitational contribution to the vertex expansion(see discussion 
at the end of section \ref{s3.1}). In addition, this analysis will shed
light on some long standing issues in SFMs such as the role of
orientation in the spin foam histories \cite{do}, the somewhat
puzzling fact that spin foam amplitudes are real rather than complex
\cite{ba-ch}, and the emergence of the cosine $\cos S_{\rm EH}$ of
the Einstein action ---rather than $e^{iS_{\rm EH}}$--- in the
classical limit \cite{limit1,limit2}.

The paper is organized as follows. In section \ref{s2} we summarize
the salient features of LQC that are needed to arrive at a sum over
histories formulation. Section \ref{s3} establishes the main results
in the timeless framework, generally used in SFMs. In particular, we
show that the physical inner product can be expressed as a vertex
expansion. In section \ref{s4} we introduce a deparametrization
using the relational time of LQC and obtain an equivalent but
\emph{distinct} vertex expansion, more directly related to the
transition amplitude. The existence of distinct vertex expansions
which sum to the same result suggests the possibility that there may
well be distinct but physically equivalent vertex amplitudes in
SFMs, each leading to a perturbative expansion that is tailored to a
specific aspect of the physical problem. To avoid repetition, we
adopted a strategy that is opposite of that used in \cite{ach1}:
here we provide detailed derivations in the timeless framework
(section \ref{s3}) and leave out the details while discussing
analogous results in the deparameterized picture (section \ref{s4}).
Section \ref{s5} summarizes the main results and discusses some
generalizations and open issues. A number of technical issues are
discussed in three Appendices.

\section{LQC: A brief overview}
\label{s2}

We will focus on the simplest LQC model that has been analyzed in
detail \cite{aps1,aps2,aps3,acs}: the k=0, $\Lambda$=0 Friedmann
model with a massless scalar field as a source. However, it should
not be difficult to extend this analysis to allow for a non-zero
cosmological constant \cite{bp,ap} or anisotropies \cite{awe2,awe3}
or to the spatially compact k=1 case \cite{apsv}.

In the FRW models, one begins by fixing a (spatial) manifold $S$,
topologically $\R^3$, Cartesian coordinates $x^i$ thereon, and a
fiducial metric $q^o_{ab}$ given by $q^o_{ab} \dd x^a \dd x^b = \dd
x_1^2 + \dd x_2^2 + \dd x_3^2$. The physical 3-metric $q_{ab}$ is
then determined by a scale factor $a$; \, $q_{ab} = a^2 q^o_{ab}$.
For the Hamiltonian analysis one fixes a cubical fiducial cell $\V$
whose volume with respect to $q^o_{ab}$ is $V_o$ so that its
physical volume is $V = a^3V_o$. The quantity $\v$ defined by $V =
2\pi\gamma \lp^2\, |\v|$ turns out to be a convenient configuration
variable, where $\gamma$ is the Barbero-Immirzi
parameter of LQG \cite{acs}.%
\footnote{In LQG the basic geometric variable is an orthonormal
triad and the physical metric $q_{ab}$ is constructed from it. If
the triad has the same orientation as the fiducial one, given by the
coordinates $x^i$, the configuration variable $\v$ is positive and
if the orientations are opposite, $\v$ is negative. Physics of the
model is insensitive to the triad orientation and hence to the sign
of $\v$. In particular the kinematic and physical quantum states
satisfy $\Psi(\v,\p) = \Psi(-\v, \p)$.}

The kinematical Hilbert space is a tensor product $\Hk=\Hkg \otimes
\Hkm$ of the gravitational and matter Hilbert spaces. Elements
$\Psi(\v)$ of $\Hkg$ are functions of $\v$ with support on a
countable number of points and with finite norm $||\Psi||^2 :=
\sum_{\v}\,|\Psi(\v)|^2$. The matter Hilbert space is the standard
one: $\Hkm = L^2(\R, \dd\p)$.
\footnote{One can also use a `polymer quantization' of the scalar
field at the kinematical level but the final physical theory turns
out to be the same.}
Thus, the kinematic quantum states of the model are functions
$\Psi(\v,\p)$ with finite norm $||\Psi||^2 := \sum_{\v}\, \sint
\dd\p\, |\Psi(\v,\p)|^2$. A (generalized) orthonormal basis in
 $\Hk$ is given by $|\v,\p\ket$ with
\be \label{kin} \bra \v^\prime,\p^\prime\,|\, \v,\p\ket =
\delta_{\v^\prime\v}\,\,\delta(\p^\prime, \p)\, .\ee
To obtain the physical Hilbert space, one first notes that the
quantum constraint can be written as
\be \label{qc} -\,\wh{C}\Psi(\v,\p) \equiv \partial^2_\p \Psi(\v,\p)
+ \Theta \Psi(\v,\p) = 0 \ee
where $\Theta$ is a positive and self-adjoint operator on $\Hkg$
\cite{kl}. More explicitly, $\Theta$ is a second order difference
operator \cite{awe2}
\ba \label{theta} \big(\Theta \Psi\big)(\v) := - \f{3\pi
G}{4\lo^2}\, \Big[\!\!&&\!\! \sqrt{|\v(\v+4\lo)|}\, (\v+2\lo)\,
\Psi(\v+4\lo)\,-\, 2\v^2\Psi(\v)\nonumber\\
&+&\, \sqrt{|\v(\v-4\lo)|}\, (\v-2\lo)\, \Psi(\v-4\lo)\, \Big]\, ,
\ea
where $\lo$ is related to the `area gap' $\Delta =4\sqrt{3}\pi
\gamma\, \lp^2$ via $\lo^2 = \,\Delta$. The form of $\Theta$ shows
that the space of solutions to the quantum constraint can be
naturally decomposed into sectors in which the wave functions have
support on specific `$\v$-lattices' \cite{aps2}. For definiteness,
we will restrict ourselves to the lattice $\v = 4n\lo$ where $n$ is
an integer. Details of the expression of $\Theta$ will not be needed
in most of our analysis.

The scalar field $\p$ is monotonic on all classical solutions (also
in the cases when k=1, and $\Lambda\not=$0) and therefore serves as
a relational time variable, a la Leibnitz, in the classical theory.
This interpretation carries over to the quantum theory. For, the
form of the quantum constraint (\ref{qc}) is similar to that of the
Klein-Gordon equation, $\p$ playing the role of time and $-\Theta$
of the spatial Laplacian (or, the elliptic operator generalizing the
Laplacian if we are in a general static space-time). Therefore, in
LQC, one can use $\p$ as an internal time variable with respect to
which physical quantities such as the density, scalar curvature,
anisotropies and shears in the Bianchi models \cite{awe2,awe3}, and 
the infinitely many modes of gravitational waves in the Gowdy models
\cite{gowdy1,gowdy2,gowdy3}, evolve.

In the spin foam literature, by contrast, one does not have access
to such a preferred time and therefore one chooses to work with the
timeless formalism. Therefore let us first forgo the emphasis on
using $\phi$ as internal time and simply implement the group
averaging procedure which uses the constraint operator as a whole,
without having to single out a preferred time variable
\cite{dm,almmt}. This procedure plays an important role in sections
\ref{s3} and \ref{s4}. Therefore it is useful to summarize it in
some detail. One begins by fixing a dense sub-space $\S$ of $\Hk$.
In LQC, this is generally taken to be the Schwartz space of smooth
functions $f(\v,\p)$ which fall off to zero at infinity faster than
any polynomial. The first step in the group averaging procedure is
to extract a solution $\Psi_f(\v,\p)$ to the quantum constraint
operator (\ref{qc}) from each $f\in \S$. These solutions are not
normalizable in $\Hk$ because the spectrum of the constraint
$\wh{C}$ on $\Hk$ is continuous. The second step of the group
averaging procedure provides an appropriate inner product between
solutions $\Psi_f(\v,\p)$.

Denote by $e_k(\v)$, with $k\in (-\infty, \infty)$ a complete set of
orthonormal eigenfunctions of $\Theta$ on $\Hkg$. We will denote the
eigenvalues by $\omega_k^2$ and, without loss of generality, assume
that $\omega_k \ge 0$ \cite{aps2,aps3}. (Eigenfunctions and operator
functions of $\Theta$ are discussed in Appendix C.) Any $f(\v,\p)
\in \S$ can be expanded as
\be f(\v,\p) =  \sint \dd k\,\,  \f{1}{2\pi}\sint\dd p_\p\,\,
\t{f}(k,p_\p)\, e^{i p_\p\, \p}\, e_k(\v)\, . \ee
\emph{Here and in what follows the range of integrals will be from
$-\infty$ to $\infty$ unless otherwise stated.} Using this
expansion, we can group-average any $f(\v,\p)$ to obtain a
distributional solution (in $\S^\star$) $\Psi_f(\v,\p)$ to the
quantum constraint:
\be \label{gpave1} \Psi_f(\v,\p) \,:=\, \sint\dd \alpha\,
[e^{i\alpha {C}}\, 2|p_\p|\, f(\v,\p)]\, =\, \sint \dd k \, \sint\dd
p_\p\,\, \delta(p_\p^2 -\omega_k^2)\,\,2|p_\p| \t{f}(k,p_\p)\,
e^{ip_\p\,\p}\, e_k(\v)\, ,\ee
where, the operator $2|p_\p|$ has been introduced just for later
technical simplification. Had we dropped it, we would have
associated with $f$ the solution $(2|p_\p|)^{-1}\, \Psi_f$ and, in
the end, obtained a unitarily equivalent representation of the
algebra of Dirac observables.

By carrying out the integral over $p_\p$ the expression of $\Psi_f$
can be brought to the desired form:
\ba \label{gpave2}\Psi_f(\v,\p) &=& \sint \dd k\,
\big[{\t{f}(k,\omega_k)}\, e^{i\omega_k\p}\,e_k(\v)\, +\, {\t{f}(k,
-\omega_k)}\, e^{-i\omega_k\p}\,e_k(\v) \big]
\nonumber\\
&=:& \Psi^+_f(\v,\p)\, +\, \Psi^-_f(\v,\p)\, . \ea
By their very definition $\Psi^\pm_f(\v,\p)$ satisfy
\be \Psi^\pm_f (\v,\p) = e^{\pm
i\sqrt{\Theta}\,(\p-\p_o)}\,\Psi^\pm_f (\v,\p_o)\, ,  \ee
whence they can be interpreted as `positive and negative frequency
solutions' to (\ref{qc}) with respect to the relational time $\p$.
Thus the group average of $f$  is a solution $\Psi_f$ to the quantum
constraint (\ref{qc}) which, furthermore, is naturally decomposed
into positive and negative frequency parts. $\Psi_f$ is to be
regarded as a distribution in $\S^\star$ which acts on elements
$g\in \S$ via the kinematic inner product \cite{dm,almmt}:
\ba (\Psi_f | g\ket &:=& \bra \Psi_f | g \ket \nonumber\\
&=& \sint \dd k \, \sint \dd p_\p\,\delta(p_\p^2-\omega_k^2)\,
2\omega_k\,\bar{\t{f}}(k,p_\p)\, \t{g}(k,p_\p)\nonumber\\
&=& \sint \dd k\, [\bar{\t{f}}(k,\omega_k)\, \t{g}(k,\omega_k) +
\bar{\t{f}}(k,-\omega_k)\, \t{g}(k,-\omega_k)] \, . \ea
Finally, the group averaged scalar product on solutions $\Psi_f$ is
given just by this action \cite{dm,almmt}. Thus, given any elements
$f,g$ in $\S$, the scalar product between the corresponding group
averaged states $\Psi_f, \Psi_g$ is given by
\be \label{pip1} (\Psi_f ,\, \Psi_g) := (\Psi_f | g\ket =
\overline{(\Psi_g | f\ket}\, .\ee
In section \ref{s3} we will obtain a vertex expansion for this
scalar product.

A conceptually important observation is that, as in the Klein-Gordon
case, there is a superselection. A complete set of Dirac observables
is given by the scalar field momentum ${p}_{\p} = -i
\partial_{\p}$ and the volume ${V}|_{\p_o}$ (or, equivalently, the
energy density operator $\rho|_{\p_o}$) at the value $\phi=\phi_o$
of the internal time. (The factor of $|p_\p|$ introduced above
simplifies the explicit expressions of ${V}|_{\p_o}$ and
$\rho|_{\p_o}$ \cite{aps2,aps3,acs}.) The action of these Dirac
observables as well as time evolution leaves the space of positive
and negative frequency solutions invariant. Therefore, as in the
Klein-Gordon theory, we are led to work with either set. In LQC, one
generally works with the positive frequency ones. Then the physical
Hilbert space $\Hp$ of LQC consists of \emph{positive frequency}
solutions $\Psi_+(\v,\p)$ to the quantum constraint (\ref{qc}), i.e.
solutions satisfying
\be\label{sch1} -i\partial_\p\,\Psi_+ (\v,\p) = \sqrt{\Theta}
\Psi_+(\v,\p)\, \equiv\, H \Psi_+(\v,\p)\ee
with inner-product (\ref{pip1}). This inner product can be
re-expressed simply as:
\be \label{pip3} (\Psi_+,\, \Phi_+)_{\rm phy} = \sum_{\v=4n\lo}\,
\bar\Psi_+ (\v,\p_o)\, \Phi_+(\v,\p_o)  \, .\ee
and is independent of the value $\p_o$ of $\p$ at which the right
side is evaluated.

While this construction of $\Hp$ does not \emph{require} us to think
of $\p$ as internal time in quantum theory, this interpretation is
natural in the light of final Eqs (\ref{sch1}) and (\ref{pip3}).
For, these equations suggest that we can think of $\v$ as the sole
configuration variable and introduce `Schr\"odinger states'
$\Psi(\v)$ through the physical inner product (\ref{pip3}). These
`evolve' via (\ref{sch1}). This is the `deparameterized' description
to which we will return in section \ref{s4}. In this picture, the
restriction to positive frequency states has direct interpretation:
${p}_{\p} \equiv \sqrt\Theta$ is now a positive operator on $\Hp$
just as ${p}_0$ is a positive operator on the traditional
Klein-Gordon Hilbert space.

\section{The timeless framework}
\label{s3}

Recall that in the spin foam literature, one works with the timeless
framework because a natural deparametrization is not available in
general. To mimic the general spin foam constructions in LQC, in
this section we will largely disregard the fact that the scalar
field can be used as relational time and that the final constraint
has the form of the Schr\"odinger equation. Instead, we will use the
group averaging procedure for the full constraint
\be {C} = -\partial_\phi^2 - \Theta \,\equiv\, {p}_{\p}^2 - \Theta
\, \ee
and incorporate the positive frequency condition in a second step.
None of the steps in this analysis refer the evolution in relational
time mentioned above. Thus, the primary object of interest will be
the physical scalar product, rather than the transition amplitude
for a Schr\"odinger state $\Psi(\v,\p_i)$ at an initial `time
instant' $\p_i$ to evolve to another state $\Phi(\v,\p_f)$ at a
final `time instant' $\p_f$.

In section \ref{s2} we considered general kinematic states
$f(\v,\p)$. In this section, by contrast, we will focus on the basis
vectors $|\v,\p\ket$ in $\Hk$ which are the LQC analogs of spin
networks that are used to specify the boundary states in SFMs.
Following the setup introduced in section \ref{s1} let us then fix
two kinematic states, $|\nu_i,\p_i \ket$ and $| \nu_f,\p_f \ket$.
For notational simplicity, we will denote the group averaged
solutions to (\ref{qc}) they define by $|[\nu_i,\p_i] \ket$ and $|
[\nu_f,\p_f] \ket$. The group averaged inner product between these
states is given by
\be \label{phy1} ([\nu_f,\p_f],\,\, [\nu_i, \p_i]) = %2\sint
%\dd{\alpha} \,\bra \nu_f,\p_f | e^{i \alpha \wh{C}} \, \sqrt\Theta|
%\nu_i, \p_i \ket =
2\sint \dd{\alpha} \, \bra \nu_f,\p_f |  \,e^{i
\alpha \wh{C}} \ |{p}_{\phi}|\,| \nu_i, \p_i \ket\, . \ee
Our goal is to express this scalar product as a vertex expansion a
la SFMs and study its properties. In section \ref{s3.1} we will
begin by rewriting it as a sum over histories a la Feynman
\cite{rpf} and then rearrange the sum as a vertex expansion. In
section \ref{s3.2} we will arrive at the same expansion using
perturbation theory in a suitably defined interaction picture. This
procedure is reminiscent of the perturbation expansion used in GFTs.
As an important consistency check, in section  \ref{s3.3} we verify
that this perturbative expansion does satisfy the constraint order
by order. Finally, in section \ref{s3.4} we observe that, in this
simple example, the coupling constant $\lambda$ used in the
expansion is intimately related to the cosmological constant
$\Lambda$. Although the precise relation we obtain is tied to LQC,
the observation illustrates in a concrete fashion how one may be
able to provide a gravitational interpretation to $\lambda$ in GFTs
and suggests an avenue for GFT to account for the smallness of
$\Lambda$.

\subsection{Sum over Histories}
\label{s3.1}

Following Reisenberger and Rovelli \cite{rr}, let us first focus on
the amplitude
\be \label{ampalpha} A(\nu_f, \p_f; \nu_i, \p_i;\alpha)= 2\, \bra
\nu_f,\p_f | \,e^{i \alpha \wh{C}} \, |{p}_{\phi}|\,| \nu_i, \p_i
\ket \, \ee
which constitutes the integrand of (\ref{phy1}). Mathematically one
can choose to regard $\alpha\wh{C}$ as a Hamiltonian operator. Then
$A(\nu_f, \p_f, \nu_i, \p_i,\alpha)$ can be interpreted as the
probability amplitude for an initial kinematic state $|\v_i,
\p_i\ket$ to evolve to a final kinematic state $|\v_f,\p_f\ket$  in
a unit `time interval' and we can follow Feynman's procedure
\cite{rpf} to express it as a sum over histories. Technically, a key
simplification comes from the fact that the constraint $\wh{C}$ is a
sum of two commuting pieces that act separately on $\Hkm$ and
$\Hkg$. Consequently, the amplitude (\ref{ampalpha}) factorizes as
\be \label{amp} A(\nu_f, \phi_f; \nu_i, \phi_i; \alpha)  = A_{\phi}
(\phi_f, \phi_i; \alpha)\, A_{G} (\nu_f, \nu_i; \alpha) \ee
with
\be A_{\phi} (\phi_f, \phi_i; \alpha) = 2\,\bra \p_f |  e^{i \alpha
p_{\p}^2 }\, |p_{\p}|  | \p_i \ket,\quad {\rm and} \quad A_{G}
(\nu_f, \nu_i; \alpha) = \bra \nu_f | e^{-i \alpha \wh{\Theta}} |
\nu_i \ket \label{ag}\, . \ee
It is easy to cast the first amplitude, $A_{\p}$, in the desired
form using either a standard Feynman expansion or simply evaluating
it by inserting a complete eigen-basis of ${p}_\p$. The result is:
\be \label{aphi} A_\p (\p_f, \p_i; \alpha)\,=\, 2\,\sint \dd p_\p \;
e^{i \alpha p_\p^2}\, e^{i p_\p (\p_f-\p_i)}\, |p_\p| \ee
The expansion of the gravitational amplitude $A_G$ is not as simple.
We will first express it as a sum over histories. In a second step,
we will evaluate the total amplitude (\ref{ampalpha}) by integrating
over $\alpha$ for each history separately. Although it is not a
priori obvious, we will find that the amplitude associated to each
history is manifestly finite and the total amplitude can be written
as a discrete sum that mimics the vertex expansion in SFMs.
%The interchange of the sum and the integral will be justified a
%posteriori.

\subsubsection{The gravitational amplitude $A_G$}
\label{s3.1.1}

As mentioned above, to apply the standard Feynman procedure we will
regard $e^{-i \alpha \wh{\Theta}}$ as an `evolution operator' with
`Hamiltonian' $\alpha \wh{\Theta}$ and a `time interval' $\Delta
\tau=1$. We emphasize that this `evolution' is a just a convenient
mathematical construct and does not correspond to the physical
evolution with respect to the relational time variables $\p$
normally used in LQC. Rather, since it is generated by the
constraint $\wh{C}$, physically it represents gauge transformations
(or time reparameterizations).

Let us divide the interval $\Delta \tau =1 $ into N parts each of
length $\epsilon = 1/N$ and write the gravitational amplitude $A_G
(\nu_f, \nu_i; \alpha)$ as
\be \bra \nu_f | e^{-i \alpha \wh{\Theta}} | \nu_i \ket =
\sum_{\bar{\v}_{N-1},...,\bar{\v}_1} \bra \v_f | e^{-i \epsilon
\alpha \wh{\Theta}} | \bar{\v}_{N-1} \ket \bra \bar{\v}_{N-1} |
e^{-i \epsilon \alpha \wh{\Theta}} | \bar{\v}_{N-2} \ket \, ... \,
\bra \bar{\v}_1 | e^{-i \epsilon \alpha \wh{\Theta}} | \v_i \ket \ee
where we have first split the exponential into N identical terms and
then introduced a decomposition of the identity operator at each
intermediate `time' $\tau= n \epsilon$, $n=1,2,..,N-1$. For
notational simplicity, we will denote the matrix element $\bra
\bar{\v}_n | e^{-i \epsilon \alpha \wh{\Theta}} | \bar{\v}_{n-1}
\ket $ by $U_{\bar{\v}_n \bar{\v}_{n-1}}$ and set $\v_f = \b\v_N$
and $\v_i = \b\v_0$. We then have
\be A_G (\nu_f, \nu_i; \alpha)= \sum_{\b\v_{N-1},\ldots
,\b\v_{1}}\,\, U_{\b\v_N\b\v_{N-1}}\, U_{\b\v_{N-1}\b\v_{N-2}}\,
\ldots \,U_{\b\v_1\b\v_0}\, . \ee
The division of $\Delta \tau$ provides a skeletonization of this
`time interval'. An assignment $\sigma_N = (\b\v_N, \ldots ,\b\v_0)$
of volumes to the $N+1$ time instants $\tau = \epsilon n$ can be
regarded as a discrete (gauge) history associated with this
skeletonization since one can envision the universe going from
$\b\v_{n-1}$ to $\b\v_n$ under a finite `evolution'. The matrix
element is given by a sum of amplitudes over these discrete
histories with fixed endpoints,
\be \label{soh1}  A_G (\v_f, \v_i; \alpha) = \sum_{\sigma_N}\,
A(\sigma_N) \equiv \sum_{\sigma_N}\,U_{\b\v_N\b\v_{N-1}}\,
U_{\b\v_{N-1}\b\v_{N-2}}\, \ldots U_{\b\v_2\b\v_1}\,U_{\b\v_1\b\v_0}
\, . \ee

The next step in a standard path integral construction is to take
the `continuum' limit, $N \to \infty$, of the skeletonization. In
particle mechanics at this stage one uses a continuous basis (say
the position basis $| x \ket$) to carry out this expansion. By
contrast, our basis $|\v_n\ket$ is discrete. As a result, one can
make rigorous sense of the  $N \to \infty$ limit by reorganizing the
well-defined sum (\ref{soh1}) according to the number of volume
transitions. The remainder of section \ref{s3.1.1} is devoted to
carrying out this step.

This task involves two key ideas. Let us first note that along a
path $\sigma_N$, the volume $\b\v$ is allowed to remain constant
along a number of time steps, then jump to another value, where it
could again remain constant for a certain number of time steps, and
so on. The first key idea is to group paths according to the number
of \emph{volume transitions} rather than time steps. Let us then
consider a path $\sigma_N^M$ which involves $M$ volume transitions
(clearly, $M\le N$):
\be \sigma_N^M = (\, \v_M,\ldots ,\v_M;\,\v_{M-1},\ldots,\v_{M-1};\,
\ldots\, \ldots ;\overbrace{\v_1,\ldots ,\v_1;\,
\underbrace{\v_0,\ldots,\v_0}_{N_1}}^{N_2}\,)\, . \ee
Thus, the volume changes from $\v_{m-1}$ to $\v_{m}$ at `time' $\tau
= N_m\epsilon$ and remains $\v_m$ till time $\tau = N_{m+1}\,
\epsilon$. Note that $\v_m$ is distinct from $\b\v_m$ used in
(\ref{soh1}): While $\v_m$ is the volume after the $m$-th
\emph{volume transition} along the given discrete path, $\b\v_m$ is
the volume at the end of the $m$-th \emph{time interval}, i.e., at
$\tau = m\epsilon$.

These discrete histories can be labeled more transparently by two
ordered sequences
\be \sigma_N^M = \{\,(\v_M, \v_{M-1},\ldots ,\v_1, \v_0);\,\, (N_M,
N_{M-1}, \ldots , N_2, N_1)\,\},\quad  \v_{m}\not=\v_{m-1},\,\, N_m
> N_{m-1}. \ee
where $\v_M,\ldots,\v_0$ denote the volumes that feature in the
history $\sigma_N^M$ and $N_k$ denotes the number of time steps
after which the volume changes from $\v_{k-1}$ to $\v_k$. Note that
while no two \emph{consecutive} volume values can be equal, a given
volume value can repeat in the sequence; $\v_{m}$ can equal some
$\v_n$ if $n\not=m\pm 1$. The probability amplitude for such a
history $\sigma_N^M$ is given by:
\be A(\sigma_N^M) = [U_{\v_M\v_M}]^{N-N_M-1}\,\, U_{\v_M
\v_{M-1}}\,\,  \ldots\,[U_{\v_1\v_1}]^{N_2-N_1-1}\,\,
U_{{\v_1}{\v_0}}\,\, [U_{\v_0\v_0}]^{N_1-1}\, . \ee

The second key idea is to perform the sum over all these amplitudes
in three steps. First we keep the ordered set of volumes $(\v_M,
\ldots , \v_0)$ fixed, but allow the volume transitions to occur at
\emph{any} value $\tau = n\epsilon$ in the interval $\Delta \tau$,
subject only to the constraint that the $m$-th transition occurs
before the ($m$+1)-th for all $m$. The sum of amplitudes over this
group of histories is given by
\be \label{AN1}  A_N(\v_M, \ldots, \v_0; \alpha) =
\sum_{N_M=M}^{N-1} \,\, \sum_{N_{M-1}=M-1}^{N_M-1}\, \ldots\,
\sum_{N_1=1}^{N_2-1}\,\, A(\sigma^M_{N}). \ee
Next we sum over all possible intermediate values of $\v_m$ such
that $\v_m\not=\v_{m-1}$, keeping $\v_0=\v_i,\, \v_M=\v_f$ to obtain
the amplitude $A_N(M)$ associated with the set of all paths in which
there are precisely $M$ volume transitions:

\be  A_N(M; \alpha) = \sum_{\substack{\v_{M-1},\ldots,\v_{1} \\
\nu_m \neq \nu_{m+1}}} \; A_N(\v_M, \ldots, \v_0; \alpha) \ee
Finally the total amplitude $A_G(\v_f;\,\v_i,\alpha)$ is obtained by
summing over all volume transitions that are permissible within our
initially fixed skeletonization with $N$ time steps:
\be \label{AG}
A_G(\v_f,\,\v_i;\alpha) = \sum_{M=0}^{N} A_N(M;\alpha) %\nonumber\\
%&\equiv \sum_{M=0}^N\,\, \Big[\sum_{\substack{\v_{M-1},\ldots,\v_{1} \\
%\nu_m \neq \nu_{m+1}}} \; A_N(\v_M,\ldots ,\v_0; \alpha)\, \Big]\, .
 \ee
This concludes the desired re-arrangement of the sum (\ref{soh1}).
The sum on the right side is manifestly finite. Furthermore, since
$A_G(\v_f,\v_i;\alpha) = \bra \nu_f | e^{-i \alpha {\Theta}} | \nu_i
\ket $, the value of the amplitude (\ref{AG}) does not depend on $N$
at all; the skeletonization was introduced just to express this
well-defined amplitude as a sum over histories. Thus, while the
range of $M$ in the sum and the amplitude $A_N(M; \alpha)$ in
(\ref{AG}) both depend on $N$, the sum does not.

Therefore we are well positioned to get rid of the skeletonization
altogether by taking the limit $N$ goes to infinity. Note first that
with our fixed skeletonization, the gravitational amplitude is a
finite sum of terms,
\be \label{sum} A_G(\v_f,\v_i;\alpha) = A_N(0; \alpha) +
A_N(1;\alpha) + \ldots + A_N(M;\alpha)+ \ldots + A_N(N;\alpha) \,\ee
each providing the contribution of all discrete paths that contain a
fixed number of volume transitions. Let us focus on the $M$th term
in the sum:
\be \label{anm} A_N(M;\alpha) = \sum_{\substack{\v_{M-1},\ldots,\v_{1}\\
\nu_m \neq \nu_{m+1}}} \; A_N(\v_M,\ldots ,\v_0; \alpha) \ee
Now, in Appendix \ref{prooflimit} we show that the limit
$\lim_{N\to\infty} A_N(\v_M,\ldots, \v_0; \alpha)$ exists and is
given by
\ba \label{lim1} A(\v_M,\ldots,\v_0; \alpha) &:=&
\lim_{N\to\infty}\, A_N(\v_M,\ldots, \v_0; \alpha)\nonumber\\
&=& \sint_0^{1} \dd\tau_M\,\, \sint_0^{\tau_M} \dd\tau_{M-1}\,\,
\ldots \sint_0^{\tau_2} \dd\tau_{1}\,\,\, A(\v_M, \ldots ,\v_0;
\tau_M,\ldots\, ,\tau_1;\alpha)\, \ea
where
\begin{align} \label{lim2}
A(\v_M, \ldots ,\v_0;\,  \tau_M,\ldots ,\tau_1;\,\alpha) :=& \,
e^{-i(1 - \tau_M) \alpha \Theta_{\v_M\v_M}}\,\,
(-i \alpha \Theta_{\v_M\v_{M-1}})\,\,
\times \nonumber\\
%& e^{-i(\tau_M-\tau_{M-1})\alpha \Theta_{\v_{M-1}\v_{M-1}}}\,
&\ldots\,\, e^{-i(\tau_2-\tau_1)\alpha \Theta_{\v_1\v_1}}\,\,
(-i \alpha \Theta_{\v_1\v_{0}})\,\, e^{-i\tau_1 \alpha
\Theta_{\v_0\v_0}}\, . \end{align}
Note that the matrix elements $\Theta_{\v_m\v_n} = \bra
\v_m|\Theta|\v_n \ket$ of $\Theta$ in $\Hkg$ can be calculated
easily from (\ref{theta}) and vanish if $(\v_m - \v_n) \not\in \{0,
\pm 4\ell_0\}$. Therefore, explicit evaluation of the limit is
rather straightforward. We will assume that the limit $N \rightarrow
\infty$ can be interchanged with the sum over $\v_{M-1}, \ldots
\v_1$. (This assumption is motivated by the fact that in the
expression of $A(\v_M, \ldots, \v_0;\alpha)$ most matrix elements of
$\Theta$ vanish, and since the initial and final volumes are fixed,
the sums over intermediate volumes $\v_{M-1}, \ldots,\v_1$ extend
over only a finite number of non-zero terms.) Then it follows that
$$A_G(M;\alpha):= \lim_{N\to\infty} A_N(M;\alpha)$$
exists for each finite $M$. Note that the reference to the
skeletonization disappears in this limit. Thus, $A_G(M;\alpha)$ is
the amplitude obtained by summing over all paths that contain
precisely $M$ volume transitions within the `time interval' $\Delta
\tau=1$, irrespective of precisely when and at what values of volume
they occurred. Finally, (\ref{sum}) implies that the total
gravitational amplitude can be written as an infinite sum:
\be \label{ag2} A_G(\v_f,\v_i; \alpha) = \sum_{M=0}^{\infty}
A_G(M;\alpha) \ee
While each partial amplitude $A_G(M; \alpha)$ is well-defined and
finite, it does not ensure that the infinite sum converges. A priori
the infinite sum on the right hand side of (\ref{ag2}) could be, for
example, only an asymptotic series to the well-defined left side.
Also, our derivation assumed that the limit $N\rightarrow \infty$
commutes with the partial sums. Both these limitations will be
overcome in section \ref{s3.2}: We will see that $A_G(\v_f,\v_i;
\alpha)$ is indeed given by a convergent sum (\ref{ag2}).

The expression (\ref{lim1}) still contains some integrals. These can
be performed exactly. The case when all of $(\v_M, \ldots ,\v_0)$
are distinct is straightforward and the result as given in
\cite{ach1}. The general case is a little more complicated and is
analyzed in Appendix \ref{gralintegral}. The final result is:

\begin{align} \label{ampG}
A(\nu_M, \ldots, \nu_0; \alpha)\, =& \, \Theta_{\v_M\v_{M-1}}
\Theta_{\v_{M-1}\v_{M-2}} \ldots \Theta_{\v_2\v_{1}} \Theta_{\v_1\v_{0}}
\,\, \times \nonumber\\
&\prod_{k=1}^p \frac{1}{(n_k-1)!} \left( \frac{\partial}{\partial
\Theta_{w_k w_k}} \right)^{n_k-1}\,\, \sum_{m=1}^p
\frac{e^{-i \alpha \Theta_{w_m w_m}\Delta \tau}}{\prod_{j \neq m}^p
( \Theta_{w_m w_m} -  \Theta_{w_j w_j})}
\end{align}
where, since the volumes can repeat along the discrete path, $w_m$
label the $p$ distinct values taken by the volume and $n_m$ the
number of times that each value occurs in the sequence.  The $n_m$
satisfy $n_1+ \ldots + n_p = M+1$.

To summarize, we have written the gravitational part $A_G (\nu_f,
\nu_i; \alpha)$ of the amplitude as a `sum over histories':
\be \label{ag3} A_G (\nu_f, \nu_i; \alpha)=\sum_{M=0}^{\infty}\,
\sum_{\substack{\v_{M-1},\ldots,\v_{1} \\
\nu_m \neq \nu_{m+1}}} \;  A(\v_M, \ldots ,\v_0; \alpha) \,
\ee
with $A(\v_M,\ldots,\v_0; \alpha)$ given by (\ref{ampG}). This
expression consists of a sum over $M$, the number of volume
transitions, and a sum over the (finite number of) sequences of
$M-1$ intermediate volumes that are consistent with the boundary
conditions and the condition that $\v_m \neq \v_{m+1}$. In section
\ref{s3.1.2} we will use this sum to generate the `vertex expansion'
of the physical inner product.

\subsubsection{Vertex expansion of the physical inner product}
\label{s3.1.2}

%In this section we combine the amplitude (\ref{aphi}) for the scalar
%field and (\ref{ag3}) for the gravitational part and carry out the
%integral over the group parameter $\alpha$ in (\ref{phy1}). This
%procedure will lead us to the desired vertex expansion.

Recall that the group-averaged scalar product can be expressed as
\be \label{phy2} ([\nu_f,\p_f],\, [\nu_i, \p_i]) = 2\sint
\dd{\alpha}\, A_\p(\v_i, \p_{i}; \alpha)\, A_G(\v_f,\v_i;\alpha)\,.
\ee
%\\
%&=& 2\sint \dd{\alpha}\, \Big[ \sint \dd p_\p \; e^{i \alpha
%p_\p^2}\, e^{i p_\p (\p_f-\p_i)}\, |p_\p|\Big]\, \Big[
%\sum_{M=0}^\infty \,\,\sum_{\substack{\v_{M-1},\ldots,\v_{1}\\ \nu_m
%\neq \nu_{m+1}}} \, A(\v_M,\ldots ,\v_0;\alpha)\, \Big]\nonumber \ea
%
The main assumption in our derivation ---the only one that will be
required also in section \ref{s3.2}--- is that one can interchange
the integration over $\alpha$ and the (convergent but infinite) sum 
over $M$ in the expression of $A_G(\v_f,\v_i;\alpha)$. Let us then 
use expressions (\ref{aphi}) and (\ref{ag3}) of $A_\p$ and $A_G$, 
make the interchange and carry out the integral over $\alpha$. The scalar product (\ref{phy2}) is then re-expressed as a sum of amplitudes associated with discrete paths $(\v_M, \ldots, \v_0)$:
\be \label{phy3} ([\nu_f,\p_f],\,\, [\nu_i, \p_i]) =
\sum_{M=0}^\infty \Big[\sum_{\substack{\v_{M-1},\ldots,\v_{1} \\
\nu_m \neq \nu_{m+1}}} \; A(\v_M,\ldots ,\v_0; \phi_f, \phi_i)\,
\Big]\, , \ee
where,
\begin{align}
\label{PhysVertExp1}
A(\v_M,\ldots ,\v_0;&\, \phi_f, \phi_i)\, =
2\, \Theta_{\v_M\v_{M-1}}\,\Theta_{\v_{M-1}\v_{M-2}}\,
\ldots\,  \Theta_{\v_2\v_{1}}\, \Theta_{\v_1\v_{0}}\,\,\times \\
\prod_{k=1}^p& \frac{1}{(n_k-1)!} \left( \frac{\partial}{\partial
\Theta_{w_k w_k}} \right)^{n_k-1}
\sum_{m=1}^p\,\, \sint \dd p_\p\, e^{ip_\p(\p_f-\p_i)}\, |p_\p|\,\,
\frac{\delta(p_\p^2- \Theta_{w_m w_m}\Delta \tau)}{\prod_{j \neq
m}^p ( \Theta_{w_m w_m} -  \Theta_{w_j w_j})}\, .\nonumber
\end{align}
The right side is a sum of distributions, integrated over $p_\p$. It
is straightforward to perform the integral and express
$A(\v_M,\ldots ,\v_0; \, \phi_f, \phi_i)$ in terms of the matrix
elements of $\Theta$:
\ba \label{physamp} A(\v_M,\ldots ,\v_0; \phi_f, \phi_i)\, &=&
\Theta_{\v_M\v_{M-1}}\,\Theta_{\v_{M-1}\v_{M-2}}\, \ldots \,
\Theta_{\v_2\v_{1}}\, \Theta_{\v_1\v_{0}}\,\, \times\\ \nonumber
&&\prod_{k=1}^p \frac{1}{(n_k-1)!} \Big(\frac{\partial}{\partial
\Theta_{w_k w_k}} \Big)^{n_k-1}\,\, \sum_{m=1}^p \frac{e^{i
\sqrt{\Theta_{w_m w_m}} \Delta \phi} + e^{-i \sqrt{\Theta_{w_m w_m}}
\Delta \phi}} {\prod_{j \neq m}^p ( \Theta_{w_m w_m} -  \Theta_{w_j
w_j})} \ea
where $\Delta \p = \p_f - \p_i$. Since by inspection each amplitude
$A(\v_M,\ldots ,\v_0, \phi_f, \phi_i)$ is real, the group averaged
scalar product (\ref{phy3}) is also real.

Finally, as explained in section \ref{s2}, the group averaging
procedure yields a solution which has both positive and negative
frequency components while the physical Hilbert space consists only
of positive frequency solutions. Let us denote the positive
frequency parts of the group averaged ket $|[\v,\p]\ket$ by
$|[\v,\p]_+\ket$. Then, the physical scalar product between these
states in $\Hp$ is given by a sum over amplitudes $A(M)$, each
associated with a fixed number of volume transitions:
\ba \label{final} ([\nu_f,\p_f]_+,\,\, [\nu_i, \p_i]_+)_{\rm phy}
&=& \sum_{M=0}^\infty \,\, A(M)\\
&=& \sum_{M=0}^\infty \,\,
\Big[\sum_{\substack{\v_{M-1},\ldots,\v_{1}\nonumber \\
\, \nu_m \neq \nu_{m+1}}} \Theta_{\v_M\v_{M-1}}\,
\Theta_{\v_{M-1}\v_{M-2}}\, \ldots \, \Theta_{\v_2\v_{1}}\,
\Theta_{\v_1\v_{0}} \\ \nonumber &\times&\,\prod_{k=1}^p
\frac{1}{(n_k-1)!} \Big(\frac{\partial}{\partial \Theta_{w_k w_k}}
\Big)^{n_k-1}\,\, \sum_{m=1}^p \frac{e^{i \sqrt{\Theta_{w_m w_m}}
\Delta \phi}} {\prod_{j \neq m}^p ( \Theta_{w_m w_m} - \Theta_{w_j
w_j})}\Big]\, . \ea
(Note that the right side is in general complex, a point to which we
will return in section \ref{s5}.) This is the vertex expansion of
the physical inner product we were seeking. It has two key features.
First, the integral over the parameter $\alpha$ was carried out and
is not divergent.  This is a non-trivial and important result if we
are interested in computing the physical inner product
perturbatively, i.e., order by order in the number of vertices.
Second, the summand involves only the matrix elements of $\Theta$
which are easy to compute. As remarked earlier, significant
simplification arises because Eq (\ref{theta}) implies that
$\Theta_{\v_m\v_n}$ is zero if $\v_m-\v_n \not\in \{0, \pm 4
\ell_0\}$.

Let us summarize. We did not begin by postulating that the physical
inner product is given by a formal path integral. Rather, we started
with the kinematical Hilbert space and the group averaging procedure
and \emph{derived} a vertex expansion of the physical inner product.
Because the Hilbert space framework is fully under control, we could
pin-point the one assumption that is needed to arrive at
(\ref{final}): the sum over vertices and the integral over $\alpha$
can be interchanged. In the full theory, one often performs formal
manipulations which result in divergent individual terms in the
series under consideration. (For instance sometimes one starts by
expanding the very first amplitude (\ref{ampalpha}) in powers of
$\alpha$ even though the $\alpha$ integral of each term is then
divergent \cite{rr,perezrev}). \emph{In our case, individual terms
in the series are all finite,} and, as we will show in section
\ref{s3.2}, even the full series (\ref{ag3}) representing the
gravitational amplitude is convergent. Nonetheless, at present the
interchange of the $\alpha$-integral and the infinite sum over $M$
has not been justified. If this gap can be filled, we would have a
fully rigorous argument that the well-defined physical inner product
admits an exact, convergent vertex expansion (\ref{final}). (This
assumption is needed only in the timeless framework because the
integration over $\alpha$ never appears in the deparameterized
framework of section \ref{s4}.) In particular, there is no need to
take a `continuum limit'.

\subsection{Perturbation Series}
\label{s3.2}

We will now show that the expression (\ref{final}) of the transition
amplitude can also be obtained using a specific perturbative
expansion. Structurally, this second derivation of the vertex
expansion is reminiscent of the perturbative strategy used in group
field theory (see, e.g., \cite{gft1,gft2}).

Let us begin by considering the diagonal and off-diagonal parts $D$
and $K$ of the operator  $\Theta$ in the basis $|\v = 4n\lo\ket$.
Thus, matrix elements of $D$ and $K$ are given by:
\be \label{DK} D_{\nu' \nu} = \Theta_{\nu \nu}\, \delta_{\nu' \nu} ,
\quad\quad K_{\nu' \nu}  =  \left\{ \begin{array}{ll} \Theta_{\nu' \nu}
& \quad \nu' \neq \nu \\
0 & \quad \nu' = \nu
\end{array}\right.
\ee
Clearly $C=p_\p^2 -D-K$. The idea is to think of $p_\p^2 - D$ as the
`main part' of $C$ and $K$ as a `perturbation'. To implement it,
introduce a 1-parameter family of operators
\be C_{\l}=p_\p^2-\Theta_{\l} :=p_\p^2 - D - \lambda K\ee
as an intermediate mathematical step. The parameter $\lambda$ will
simply serve as a marker to keep track of powers of $K$ in the
perturbative expansion and we will have to set $\lambda=1$ at the
end of the calculation.

Our starting point is again the decomposition (\ref{amp}) of the
amplitude $A(\v_f,\p_f; \v_i,\p_i; \alpha)$ into a scalar field and
a gravitational part. The $\l$ dependance appears in the
gravitational part:
\be A^{(\l)}_{G} (\nu_f, \nu_i, \alpha) := \bra \nu_f | e^{-i \alpha
\Theta_\l} | \nu_i \ket . \ee
Let us construct a perturbative expansion of this amplitude. Again we
think of $e^{-i \alpha \Theta_\l} $ as a mathematical `evolution
operator' defined by the `Hamiltonian' $\alpha \Theta_\l$  and a
`time interval' $\Delta \tau =1$. The ` unperturbed Hamiltonian' is
$\alpha D$ and the `perturbation' is $\l \alpha K$. Following the
textbook procedure, let us define the `interaction Hamiltonian' as
\be H_I(\tau)=e^{i \alpha D \tau}\,\, \alpha K \,\, e^{-i \alpha D
\tau }. \ee
Then the evolution in the interaction picture is dictated by the
1-parameter family of unitary operators on $\Hkg$
\be \t{U}_\lambda(\tau)= e^{i \alpha D\tau} e^{-i \alpha
\Theta_{\l}\,\tau}\, , \quad\quad {\rm satisfying} \quad
\f{\dd\t{U}_\l(\tau)}{\dd \tau} = - i\l\,
H_I(\tau)\t{U}_\lambda(\tau)\, . \ee
The solution of this equation is given by a time-ordered
exponential:
\ba  \t{U}_\lambda(\tau) & = & \mathcal{T}\,\,
e^{-i \sint_0^{\tau} H_I(\tau)\dd\tau}\nonumber \\
& = & \sum_{M=0}^{\infty}\,\,\l^M\, \sint_{0}^{\tau} \dd\tau_{M}
\,\sint_{0}^{\tau_{M}} \dd\tau_{M-1}\ldots \sint_{0}^{\tau_2}
\dd\tau_1 \,\,\,  [-iH_I(\tau_{M})]\, ...\, [-iH_I(\tau_{1})]\,
.\label{utilde} \ea

Next we use the relation  $e^{-i \alpha \Theta_\l} = e^{-i \alpha
D}\t{U}_\lambda(1)$, with $\t{U}_\lambda$ given by (\ref{utilde}),
take the matrix element of $e^{i\alpha \Theta_{\l}}$ between initial
and final states, $|\nu_i \equiv \nu_0\ket$ and $|\nu_f \equiv
\nu_M\ket$, and write out explicitly the product of the $H_I$'s. The
result is
\ba A^{(\l)}_{G} (\nu_f, \nu_i, \alpha)  = \sum_{M=0}^{\infty}\,\,
\l^M\, \sint_{0}^{1}\,\, \dd\tau_{M}\!\!&...&\!\!\sint_{0}^{\tau_2}
\dd\tau_1\,\, \sum_{\v_{M-1},\,\ldots,\,\v_{1}} [e^{-i
(1-\tau_M) \alpha D_{\nu_M \nu_M}}]\,\,\times\nonumber\\
&& \,\,(-i \alpha K_{\nu_M\nu_{M-1}})\,\,\ldots\, (-i \alpha
K_{\v_1\v_0})\,\, [e^{-i\tau_1 \alpha D_{\v_0\v_0}}]\, . \ea
We can now replace $D$ and $K$ by their definition (\ref{DK}).
Because $K$ has no diagonal matrix elements, only the terms with
$\v_m\not=\v_{m+1}$ contribute and the sum reduces precisely to
\be \label{alambda} A^{(\l)}_{G} (\v_f, \v_i, \alpha) =
\sum_{M=0}^{\infty}\, \lambda^M \,
\Big[\sum_{\substack{\v_{M-1},\ldots,\v_{1} \\
\v_m \neq \v_{m+1}}} \, A(\v_M,\ldots, \v_0;\alpha)\,\Big]\, , \ee
where $A(\v_M,\ldots,\v_0;\alpha)$ is given by  (\ref{ampG}) as in
the sum over histories expansion of section \ref{s3.1.1}.

We can now construct the total amplitude by including the scalar
field factor (\ref{aphi}) and performing the $\alpha$ integral as in
section \ref{s3.1.2}. %(\ref{PhysVertExp1}.)
Then the group averaged scalar product is given by
\be \label{phy4} ([\nu_f,\p_f],\, [\nu_i, \p_i] )^{(\l)}
= \sum_{M=0}^\infty \l^M \Big[\sum_{\substack{\v_{M-1},\ldots,\v_{1} \\
\nu_m \neq \nu_{m+1}}} \; A(\v_M,\ldots ,\v_0, \phi_f, \phi_i)\, \Big]\,
\ee
where $A(\v_M,\ldots ,\v_0, \phi_f, \phi_i)$ is given in
(\ref{physamp}). If we now set $\l=1$, (\ref{phy4}) reduces to
(\ref{phy3}) obtained independently in section \ref{s3.1.2}.

Finally, let us restrict ourselves to the positive frequency parts
$|[\v, \p]_+\ket$ of $[\v,\p]\ket$ which provide elements of $\Hp$.
Reasoning of section \ref{s3.1.2} tells us that the physical scalar
product $([\v_f,\p_f]_+,\, [v_i,\p_i]_+)_{\rm phy}$ is given by
(\ref{final}).

Thus, by formally regarding the \emph{volume changing}, off-diagonal
piece of the constraint as a perturbation we have obtained an
independent derivation of the vertex expansion for
$([\v_f,\p_f]_+,\, [v_i,\p_i]_+)_{\rm phy}$ as a power series
expansion in $\lambda$, the power of $\lambda$ serving as a bookmark
that keeps track of the number of vertices in each term. In this
sense this alternate derivation is analogous to the vertex expansion
obtained using group field theory. This derivation has a technical
advantage. Since $H_I$ is self-adjoint on $\Hkg$, it follows that
the expansion (\ref{utilde}) of $\t{U}_{\lambda}(\tau)$ is
convergent everywhere on $\Hkg$ \cite{rs}. This in turn implies that
the right hand side of (\ref{alambda}) converges to the well-defined
gravitational amplitude $A^{(\l)}_{G} = \bra \v_f| e^{-i\alpha
\Theta_{\l}}|\v_i\ket$. However, to arrive at the final vertex
expansion starting from (\ref{alambda}) we followed the same
procedure as in section \ref{s3.1.2}. Therefore, this second
derivation of the vertex amplitude also assumes that one can
interchange the integral over $\alpha$ with the (convergent but)
infinite sum over $M$ in (\ref{alambda}).

\subsection{Satisfaction of the constraint}
\label{s3.3}

The physical inner product between the basis states defines a
2-point function:
\be G(\v_f,\p_f;\,\v_i\p_i) := ([\v_f,\p_f]_+,\, [\v_i,\p_i]_+)_{\rm
phy} \ee
and it follows from section \ref{s2} that it satisfies the
constraint equation in each argument. Since $G(\v_f,\p_f;\,\v_i\p_i)
=\bar{G}(\v_i,\p_i;\,\v_f\p_f)$, it suffices to focus just on one
argument, say the final one. Then we have:
\be \label{Geq} [\partial^2_{\p_f} - \Theta_f]G(\v_f,\p_f;\,\v_i,
\p_i) =0\ee
where $\Theta_f$ acts as in (\ref{theta}) but on $\v_f$ in place of
$\v$. If one replaces $\Theta$ by $\Theta_\l$, one obtains a 2-point
function $G_\l(\v_f,\p_f;\, v_i,\p_i)$ which, as we saw in section
\ref{s3.2} admits a perturbative expansion:
\be \label{G1} G_\l(\v_f,\p_f;\, v_i,p_i) = \sum_{M=0}^\infty\,
\lambda^M\, A_M(\v_f,\p_f; v_i,\p_i), \ee
where $A_M$ is the amplitude defined in (\ref{final}):
\ba \label{G2} A_M (\v_f,\p_f; v_i,\p_i)&=&
\sum_{\substack{\v_{M-1},\ldots,\v_{1} \\
\, \nu_m \neq \nu_{m+1}}}\, A_+(\v_M,\ldots\v_0;\,
\p_f,\p_i)\nonumber\\
&\equiv& \sum_{\substack{\v_{M-1},\ldots,\v_{1}\nonumber \\
\, \nu_m \neq \nu_{m+1}}} \Theta_{\v_M\v_{M-1}}\,
\Theta_{\v_{M-1}\v_{M-2}}\, \ldots \, \Theta_{\v_2\v_{1}}\,
\Theta_{\v_1\v_{0}} \, \times\nonumber\\ &\, &\prod_{k=1}^p
\frac{1}{(n_k-1)!} \Big(\frac{\partial}{\partial \Theta_{w_k w_k}}
\Big)^{n_k-1}\,\, \sum_{m=1}^p \frac{e^{i \sqrt{\Theta_{w_m w_m}}
\Delta \phi}} {\prod_{j \neq m}^p ( \Theta_{w_m w_m} - \Theta_{w_j
w_j})}\ea
The suffix $+$ in $A_+(\v_M,\ldots, \v_0;\,\p_f,\p_i)$ emphasizes
that we have taken the positive frequency part.

As a non-trivial check on this expansion we will now show that
$G_\l$ satisfies (\ref{Geq}) order by order. Since $\Theta_\l = D +
\l K$, our task reduces to showing
\be \label{Aeq} (\partial^2_{\p_f} - D_f)\,A_M(\v_f,\p_f;
\,\v_i\p_i) - K_f\, A_{M-1}(\v_f,\p_f;\,\v_i\p_i) =0 \, . \ee
We will show that the left hand side is zero path by path in the sense
that for every path acted on by the off-diagonal part there are two
paths acted on the diagonal part that cancel it. Without loss of
generality we assume that $\v_f = w_p$ in (\ref{G2}). Then we have
\begin{align}
(\partial_{\p_f}^2 - {D_f})  A_+(\nu_f, \nu_{M-1}, \ldots,
\v_1, \v_i; \phi_f, \phi_i) = \Theta_{\v_f\v_{M-1}}
\Theta_{\v_{M-1}\v_{M-2}} \ldots  \Theta_{\v_2\v_{1}}
\Theta_{\v_1\v_i} \times \nonumber\\
\left[ \prod_{k=1}^p \frac{1}{(n_k-1)!} \left(
\frac{\partial}{\partial  \Theta_{w_k w_k}} \right)^{n_k-1}
\sum_{m=1}^p \frac{ \Theta_{w_m w_m}e^{i \sqrt{\Theta_{w_m w_m}}
\Delta \phi}}{\prod_{j \neq m}^p ( \Theta_{w_m w_m} -
\Theta_{w_j w_j})} \right. \\ \nonumber -
\left. \Theta_{w_p w_p} \prod_{k=1}^p \frac{1}{(n_k-1)!}
\left( \frac{\partial}{\partial  \Theta_{w_k w_k}}
\right)^{n_k-1}\,\, \sum_{m=1}^p \frac{e^{i \sqrt{\Theta_{w_m w_m}}
\Delta \phi}}{\prod_{j \neq m}^p ( \Theta_{w_m w_m} -
\Theta_{w_j w_j})} \right]\, . \end{align}
If $w_p$ occurs with multiplicity $n_p=1$, if $\v_f$ is the only
volume to take the value $w_p$ then there are no derivatives in
$\Theta_{w_pw_p}$ in the above equation and it simplifies to
\begin{align}
(\partial_{\p_f}^2 - {D_f})  A_+(\nu_f, \nu_{M-1}, \ldots,
\v_1, \v_i; \phi_f, \phi_i) = \Theta_{\v_f\v_{M-1}}
\Theta_{\v_{M-1}\v_{M-2}} \ldots  \Theta_{\v_2\v_{1}}
\Theta_{\v_1\v_i} \times \nonumber \\
\left[ \prod_{k=1}^{p-1} \frac{1}{(n_k-1)!}
\left( \frac{\partial}{\partial  \Theta_{w_k w_k}} \right)^{n_k-1}\,\,
\sum_{m=1}^{p} \frac{ (\Theta_{w_m w_m}- \Theta_{w_p w_p})
e^{i \sqrt{\Theta_{w_i w_i}} \Delta \phi}}{\prod_{j \neq i}^{p}
(\Theta_{w_m w_m} -  \Theta_{w_j w_j})} \right] \nonumber \\
{}\nonumber\\
= \Theta_{\v_f\v_{M-1}} A_+(\nu_{M-1},
\ldots, \v_1, \v_i; \phi_f, \phi_i)\, .
\end{align}
Thus, on simple paths where the final volume occurs only once in the
sequence, the action of $[\partial_{\p_f}^2 - {D}]$ is to give the
amplitude of the path without $\v_f$, times a matrix element of
$\Theta$ related to the transition from $\v_{M-1}$ to $\v_f$. In
general, the value of the final volume can be repeated in the
discrete path; $n_p \neq 1$. In that case we need to push
$\Theta_{w_p w_p}$ under the derivatives but the final result is the
same. Thus, in all cases we have
%
% \begin{align}
%(\wh{p_\p}^2 - \wh{D})  A(\nu_f, \nu_{M-1}, \ldots, \v_1, \v_i, \phi_f, \phi_i)
%= \Theta_{\v_f\v_{M-1}}\Theta_{\v_{M-1}\v_{M-2}} \ldots
%\Theta_{\v_2\v_{1}} \Theta_{\v_1\v_i} \times \\
%\nonumber  \left[ \prod_{k=1}^p \frac{1}{(n_k-1)!}
%\left( \frac{\partial}{\partial  \Theta_{w_k w_k}}
%\right)^{n_k-1} \sum_{i=1}^p \frac{(\Theta_{w_i w_i}-
%\Theta_{w_p w_p}) e^{i \sqrt{\Theta_{w_i w_i}} \Delta \phi}}
%{\prod_{j \neq i}^p ( \Theta_{w_i w_i} -  \Theta_{w_j w_j})} \right. \\
%+  \left. \prod_{k=1}^{p-1} \frac{1}{(n_k-1)!} \left( \frac{\partial}{\partial
%\Theta_{w_k w_k}} \right)^{n_k-1} \frac{1}{(n_p-2)!}
%\left( \frac{\partial}{\partial  \Theta_{w_p w_p}}
%\right)^{n_p-2}\sum_{i=1}^p \frac{e^{i \sqrt{\Theta_{w_i w_i}}
%\Delta \phi}}{\prod_{j \neq i}^p ( \Theta_{w_i w_i} -
%\Theta_{w_j w_j})} \right]
 %\end{align}
 %This simplifies down to
%
\be (\partial_{\p_f}^2 - {D_f})\, A_+(\v_f, \v_{M-1}, \ldots, \v_1,
\v_i; \phi_f, \phi_i) = \Theta_{\v_f\v_{M-1}}\, A_+(\nu_{M-1},
\ldots, \v_1, \v_i; \phi_f, \phi_i)\, . \ee
Finally, it is straightforward to evaluate the action of the
off-diagonal part on $A_{M-1}$ (see (\ref{Aeq})):
\be {K}\, A_+(\v_f, \v_{M-2}, \ldots, \v_1, \v_i; \phi_f, \phi_i) =
\sum_{\v_{M-1}} \Theta_{\v_f\,\v_{M-1}} A_+(\v_{M-1}, \v_{M-2},
\ldots, \v_1, \v_i; \phi_f, \phi_i)\, . \ee

Combining these results we see that Eq. (\ref{Aeq}) is satisfied.
Thus the vertex expansion we obtained is a solution to the quantum
constraint equation.  Further it is a good perturbative solution in
the sense that, if we only take paths in which the number of volume
transitions is less than some ${M}^\star$, then the constraint is
satisfied to the order $\l^{{M}^\star}$:
\be [\partial^2_{\p_f} - (D_f + \l K_f)]\,\sum_{M=0}^{{M}^\star}\,
\l^M\, A_M(\v_f,\p_f; \,\v_i,\p_i) = \mathcal{O} (\l^{{M}^\star+1})
\ee
Also in this calculation the cancelations occur in a simple manner;
the off-diagonal part acting on paths with $M-1$ transitions gives a
contribution for each path with $M$ transitions that could be
obtained by a adding a single additional transition in the original
path. These contributions cancel with the action of the diagonal
part on the paths with $M$ transitions.

This calculation provides an explicit check on our perturbative
expansion of the physical inner product. This is a concrete
realization, in this simple example, of a central hope of SFMs: to
show that the physical inner product between spin networks,
expressed as a vertex expansion, does solve the Hamiltonian
constraint of LQG order by order.

\subsection{The `coupling constant' $\lambda$ and the cosmological
constant $\Lambda$}
\label{s3.4}

So far we have regarded the GFT inspired perturbation theory as a
calculational tool and the coupling constant $\lambda$ as a
book-keeping device which merely keeps track of the number of
vertices in the vertex expansion. From this standpoint values of
$\lambda$ other than $\lambda=1$ have no physical significance.
However, if one regards GFT as fundamental and gravity as an
emergent phenomenon, one is forced to change the viewpoint. From
this new perspective, the coupling constant $\l$ is physical and
can, for example, run under a renormalization group flow. The
question we raised in section \ref{s1} is: What would then be the
physical meaning of $\lambda$ from the \emph{gravitational}
perspective? Surprisingly, in the LQC model under consideration,
$\lambda$ can be regarded as (a function of) the cosmological
constant $\Lambda$.

Let us begin by noting how the quantum constraint changes in
presence of a cosmological constant $\Lambda$:
\be - C(\Lambda) \,=\, \partial_\p^2 + \Theta(\Lambda)\, \equiv \,
\partial_\p^2 + \Theta - \pi G \gamma^2 \Lambda \v^2\, . \ee
Thus, only the diagonal part of $\Theta$ is modified and it just
acquires an additional term proportional to $\Lambda$. In the
GFT-like perturbation expansion, then, we are led to decompose
$\Theta_\lambda(\Lambda)$ as
\be \Theta_{\lambda}(\Lambda) = D(\Lambda) + \lambda K \qquad {\rm
where} \qquad D(\Lambda)  = \pi G \,(\f{3}{2\lo^2} -
\gamma^2\Lambda)\, \v^2\, .\ee
It is now easy to check that $\Psi(\v,\p)$ satisfies the constraint
equation
\be[\partial_\p^2 +D(\Lambda) + \lambda K]\, \Psi(\v,\p) =0\, \ee
with cosmological constant $\Lambda$ if and only if
$\t{\Psi}(\v,\t\p)$ satisfies
\be \label{Lambdatilde}
[\partial_{\t{\p}}^2 + D(\t{\Lambda}) + K]\, \t{\Psi}(\v,\t{\p})
= 0\ee
where
\be \t{\Lambda} = \f{\Lambda}{\l} +
\f{3}{2\gamma^2\lo^2\l}\, (\l-1), \qquad \t{\p} =
\sqrt{\l}\, {\p}, \quad {\rm and}\quad \t{\Psi}(\v,\t{\p}) =
\Psi(\v,\p)\, . \ee
Consequently the two theories are isomorphic. Because of this isomorphism, the gravitational meaning of the coupling constant 
$\l$ is surprisingly simple: it is related to the cosmological 
constant $\Lambda$.

Suppose we want to consider the Hamiltonian theory (or the SFM) for 
zero cosmological constant. Then we are interested in the Hamiltonian constraint (\ref{Lambdatilde}) with $\t\Lambda=0$. From the GFT 
perspective, on the other hand, the cosmological constant is $\Lambda$ 
which `runs with the coupling constant' $\lambda$ via%
\footnote{Note incidentally that, contrary to what is often assumed, running of constant under a renormalization group flow is not related 
to the physical time evolution in cosmology \cite{rw}.}
\be \Lambda = \f{3}{2\gamma^2\lo^2} (1- \lambda) \ee  
At $\l=1$, we have $\Lambda=0$, whence the GFT reproduces the amplitudes of the SFM with zero cosmological constant. The question is: What is the space-time interpretation of GFT for other values of $\lambda$? From the perturbation theory perspective, $\lambda$ will start out being zero in GFT and, under the renormalization group flow, it will hopefully increase to the desired value $\lambda=1$. In the \emph{weak coupling limit} $\lambda \approx 0$, the SFM will reproduce the amplitudes of the theory which has a positive but Planck scale cosmological constant $\Lambda \approx 3/2\gamma^2\lo^2$. This is just what one would expect from the `vacuum energy' considerations in quantum field theories in Minkowski space-time. As the coupling constant $\lambda$ increases and approaches the SFM value $\lambda=1$, the cosmological constant $\Lambda$ decreases. Now, suppose that the renormalization group flow leads us close to but not all the way to $\lambda=1$. If we are just slightly away from the fixed point $\lambda=1$, the cosmological constant $\Lambda$ would be small and positive. These considerations are only heuristic. But they suggest an avenue by which a fully developed GFT could perhaps account for the smallness of the cosmological constant.

\section{Deparameterized Framework}
\label{s4}

In this section we will use the deparameterized framework which
emphasizes the role of $\p$ as internal time. As explained in
section \ref{s2}, now we can work in the Schr\"odinger picture,
regarding $\v$ as the configuration variable and $\p$ as time. The
physical states are now represented as functions $\Psi(\v)$ with a
finite norm,
\be \label{norm}||\Psi||^2_{\rm phy} = \sum_{\v=4n\lo}\,
|\Psi(\v)|^2\, ,\ee
and they evolve via Schr\"odinger equation:
\be\label{sch2} -i\partial_\p\,\Psi (\v,\p) = \sqrt{\Theta}
\Psi(\v,\p)\, \equiv\, H \Psi(\v,\p)\, .\ee
In contrast to section \ref{s3}, in this section we will not be
interested in the kinematical Hilbert space or the group averaging
procedure. The primary object of interest will rather be the
\emph{transition amplitude}
\be A(\v_f,\vp;\v_i,0) = \bra \v_f|\, e^{iH\vp}|\v_i\ket \ee
for the initial physical state $|\v_i\ket$ at time $\p_i =0$ to
evolve to $|\v_f\ket$ at time $\p_f = \vp$. From our discussion in
section \ref{s2}, one would expect this amplitude to equal the
physical scalar product $([\v_f,\vp]_+,\, [\v_i,0]_+)_{\rm phy} =
G(\v_f,\vp;\v_i,0)$ considered in section \ref{s3}. This is indeed
the case. For, the positive frequency solution
$\Psi_{\v_i,\p_i} \equiv [\v_i,\p_i]_+$ obtained by group averaging
the kinematic basis vector $|\v_i,\p_i\ket$ is given by
\be \Psi_{\v_i,\p_i}(\v,\p) = \sint \dd k\, (\bar{e}_k(\v_i)\,
e^{-i\w_k\p_i})\, e^{i\w_k(\p)}\,  e_k(\v) \ee
(see Eq.(\ref{gpave2})) so that the physical scalar product between
positive frequency solutions $[\v_i,\p_i]_+$ and $[\v_f,\p_f]_+$ is
given by
\be ([\v_f,\p_f]_+ ,\,[\v_i,\p_i]_+)_{\rm phy} = \sint \dd k\,
e^{i\w_k(\p_f-\p_i)}\,\, \bar{e}_k(\v_i)\,e_k(v_f) \ee
(see Eq (\ref{pip1})). The right hand side is precisely the
expression of the transition amplitude $\bra \v_f|\,
e^{iH\vp}|\v_i\ket = \sint \dd k\, \bra \v_f|\, e^{iH\vp}|k\ket \bra
k|\v_i\ket$. Since $e_k(\v) = \bra\v|k\ket$, we have the equality:
$G(\v_f,\vp;\v_i,0) = A(\v_f,\vp;\v_i,0)$. However, the
interpretation now emphasizes the \emph{physical} time-evolution in
$\p$ generated by $H$ whence $A(\v_f,\vp;\v_i,0)$ has the
interpretation of a physical transition amplitude. Therefore, we can
literally follow ---not just mimic--- the procedure Feynman used in
non-relativistic quantum mechanics \cite{rpf}. This will again lead
to a vertex expansion but one which, if terminated at any finite
order, is distinct from that obtained in section \ref{s3}.

In spite of important conceptual differences, the mathematical
procedure used in this section is completely analogous to that used
in section \ref{s3}. Furthermore, this deparameterized framework was
discussed in greater detail than the timeless framework in
\cite{ach1}. Therefore, in this section we will present only the
main steps.

\subsection{Sum over histories}
\label{s4.1}

Following Feynman, let us divide the time interval $(\vp,0)$ into
$N$ equal parts, each of length $\epsilon= \vp/N$, and express the
transition amplitude $A(\v_f,\vp;\v_i,0)$ as a sum over discretized
paths $\sigma_N = (\v_f=\b\v_N, \b\v_{N-1}, \ldots, \b\v_1, \b\v_0
=\v_i)$:
\be \label{soh2} A(\v_f,\varphi;\,\v_i,0) = \sum_{\sigma_N}\,
A(\sigma_N)\quad\quad {\rm with}\quad A(\sigma_N) =
U_{\b\v_N\b\v_{N-1}}\, U_{\b\v_{N-1}\b\v_{N-2}}\, \ldots
U_{\b\v_2\b\v_1}\,U_{\b\v_1\b\v_0} \,  \ee
where now  $U_{\b\v_{n+1}\b\v_n} \equiv \bra \b\v_{n+1}|e^{i
\epsilon H } |\b\v_n\ket $. The structure of Eq (\ref{soh2})
parallels that of Eq (\ref{soh1}) in section \ref{s3.1}. However,
the mathematical `time interval' $\Delta \tau =1$ in section
\ref{s3.1} is now replaced by the \emph{physical} time interval
$(\vp,0)$ and the mathematical `Hamiltonian' $\alpha \Theta$ by the
\emph{physical} Hamiltonian $H = \sqrt{\Theta}$. Furthermore we no
longer split the amplitude into a gravitational part and a scalar
field part and the group averaging parameter $\alpha$ will never
appear in this section.

As in section \ref{s3.1}, the next step is to make a convenient
rearrangement of this sum, emphasizing volume-transitions, rather
than what happens at each point $\p_n= n\epsilon$ of the
skeletonized time interval. Thus, we first recognize that the volume
could remain constant for a number of time steps and consider
histories $\sigma_N^M$ with precisely $M$ volume transitions (where
$M<N$):
\be \sigma_N^M = \{\,(\v_M, \v_{M-1},\ldots ,\v_1, \v_0);\,\, (N_M,
N_{M-1}, \ldots , N_2, N_1)\,\},\quad  \v_{m}\not=\v_{m-1},\,\, N_m
> N_{m-1}. \ee
where $\v_M,\ldots,\v_0$ denote the volumes that feature in the
history $\sigma_N^M$ and $N_k$ denotes the number of time steps
after which the volume changes from $\v_{k-1}$ to $\v_k$. The
probability amplitude for such a history $\sigma_N^M$ is given by:
\be A(\sigma_N^M) = [U_{\v_M\v_M}]^{N-N_M-1}\,\, U_{\v_M
\v_{M-1}}\,\, \ldots\, [U_{\v_1\v_1}]^{N_2-N_1-1}\,\,
U_{{\v_1}{\v_0}}\,\, [U_{\v_0\v_0}]^{N_1-1}\, . \ee
As in section \ref{s3.1}, we carry out the sum over all these
amplitudes in three steps. First we keep the ordered set of volumes
$(\v_M, \ldots , \v_0)$ fixed, but allow the volume transitions to
occur at \emph{any} value $\p = n\epsilon$ in the interval $\I$,
subject only to the constraint that the $m$-th transition occurs
before the ($m$+1)-th for all $m$. The sum of amplitudes over this
group of histories is given by
\be \label{1} A_N(\v_M, \ldots, \v_0) = \sum_{N_M=M}^{N-1} \,\,
\sum_{N_{M-1}=M-1}^{N_M-1}\, \ldots\, \sum_{N_1=1}^{N_2-1}\,\,
A(\sigma^M_{N})\, . \ee
Next we sum over all possible intermediate values of $\v_m$ such
that $\v_m\not=\v_{m-1}$, keeping $\v_0=\v_i,\, \v_M=\v_f$, to
obtain the amplitude $A(M)$ associated with the set of all paths in
which there are precisely $M$ volume transitions:
\be \label{2} A_N(M) = \sum_{\substack{\v_{M-1},\ldots,\v_{1} \\
\nu_m \neq \nu_{m+1}}} \; A_N(\v_M, \ldots, \v_0) \ee
Finally the total amplitude $A(\v_f,\p;\,\v_i,0)$ is obtained by
summing over all volume transitions that are permissible within our
initially fixed skeletonization with $N$ time steps:
\be \label{3} A(\v_f,\varphi;\,\v_i,0) = \sum_{M=0}^{N} A_N(M)
\equiv \sum_{M=0}^N\,\, \Big[\sum_{\substack{\v_{M-1},\ldots,\v_{1} \\
\nu_m \neq \nu_{m+1}}} \; A_N(\v_M,\ldots ,\v_0)\, \Big]\, .
 \ee

As in section \ref{s3.1}, since $A(\v_f,\varphi;\,\v_i,0) =
\bra\v_f|e^{iH\varphi}|\v_i\ket$, the value of the amplitude
(\ref{3}) does not depend on $N$ at all; the skeletonization was
introduced just to express this well-defined amplitude as a sum over
histories. Thus, while the range of $M$ in the sum and the amplitude
$A_N(M)$ in (\ref{3}) both depend on $N$, the sum does not. We can
get rid of the skeletonization altogether by taking the limit as $N$
goes to infinity, to express the total transition amplitude as a
vertex expansion in the spirit of the timeless framework of
spin-foams. Reasoning analogous to that in Appendix A shows that the
limit does exist. In this limit the reference to the skeletonization
of the time interval disappears and volume changes can now occur at
\emph{any} time in the continuous interval $(\p_i=0,\, \p_f=\vp)$.
The contribution $A_M$ from paths with precisely $M$ volume changes
has a well defined `continuous time' limit and the total amplitude
is given by a discrete sum over $M$:
\be \label{finalamp} A(\v_f,\vp;\, \v_i,0) = \sum_{M=0}^\infty\,
A_M(\v_f,\vp;\, \v_i,0) \ee
where the partial amplitudes $A_M$ are given by
\ba \label{partialamp} A_M (\v_f,\vp;\, \v_i,0)&=&
\sum_{\substack{\v_{M-1},\ldots,\v_{1} \\  \nu_m \neq
\nu_{m+1}}} A(\v_f, \v_{M-1}, \ldots \v_1, \v_i, \varphi)\\
&=& \sum_{\substack{\v_{M-1},\ldots,\v_{1} \\  \nu_m \neq
\nu_{m+1}}} H_{\v_M\v_{M-1}} H_{\v_{M-1}\v_{M-2}} \ldots
H_{\v_2\v_{1}} H_{\v_1\v_{0}}\, \times \nonumber\\
&&\prod_{k=1}^p \frac{1}{(n_k-1)!} \left( \frac{\partial}{\partial
H_{w_k w_k}} \right)^{n_k-1} \sum_{m=1}^p \frac{e^{i H_{w_m w_m}
\varphi}}{\prod_{j \neq m}^p ( H_{w_m w_m} -  H_{w_j w_j})}\, .
\nonumber\ea
As one might expect, the final expression involves just the matrix
elements of the Hamiltonian $H= \sqrt{\Theta}$. These are calculated
in Appendix \ref{matrixel}.

Thus, the total transition amplitude has been expressed as a vertex
expansion (\ref{finalamp}) a la SFMs. We provided several
intermediate steps because, although the left hand sides are equal,
the final vertex expansions is \emph{different} from that obtained
in section \ref{s3.1}: While (\ref{finalamp}) features matrix
elements of $H= \sqrt{\Theta}$, (\ref{final}) features matrix
elements of $\Theta$ itself. The existence of distinct but
equivalent vertex expansions is quite surprising. In each case we
emphasized a distinct aspect of dynamics: the timeless framework and
group averaging in (\ref{final}), and relational time and
deparametrization in (\ref{finalamp}).

\subsection{Perturbation expansion}
\label{s4.2}

This vertex expansion can also be obtained as a perturbation series
that mimics GFTs. As in section \ref{s3}, the perturbative approach
avoids skeletonization altogether and has the advantage that it
guarantees a convergent series. Furthermore, since this
deparametrization approach does not refer to an integral over
$\alpha$, the assumption of interchange of the integral and the sum
over $M$ that was required in section \ref{s3.2} is no longer
necessary.

Let us now focus on the Hamiltonian operator $H = \sqrt{\Theta}$
(rather than on $\Theta$ used in section \ref{s3.2}) and decompose
it into a diagonal part $D$ and the remainder, non-diagonal part $K$
which is responsible for a volume change. Finally, let us set $H_\l
= D + \l K$ where $\l$ will serve as a marker for powers of $K$,
i.e., the number of volume changes in the expansion. Then, by
working in the appropriate interaction picture, we obtain:
\be \label{fullamp2} A_\l(\v_f, \varphi ; \v_i, 0) =
\sum_{M=0}^{\infty} \l^M A_M (\v_f, \varphi ; \v_i, 0) \ee
where $A_M$ is again given by (\ref{partialamp}). This power series
in $\l$ is reminiscent of what one finds in GFTs. If we set $\l=1$
at the end of this derivation, we recover the vertex expansion
(\ref{finalamp}) a la SFMs. For a discussion of the intermediate
steps, see \cite{ach1} and Appendix A.

\subsection{Satisfaction of the Schrodinger Equation}
\label{s4.3}

Recall that in the deparametrization scheme, the Schr\"odinger
equation (\ref{sch2}) incorporates both the quantum constraint and
the positive frequency condition. By its very definition, the exact
transition amplitude $A(\v_f,\vp; v_i,0)$ satisfies this
Schr\"odinger equation. As a check on the perturbative expansion
(\ref{fullamp2}) we are led to ask whether the Schr\"odinger
equation would be satisfied in a well-controlled approximate sense
if we were to truncate the series on the right side of
(\ref{fullamp2}) at a finite value, say ${M}^\star$ of $M$. We will
now show that this is indeed the case.

Since $H_\l = D + \l K$, the schr\"odinger equation would be solved
order by order in perturbation series if for each $M$ we have:
\be (i\partial_{\vp} + D_f )\, A_M(\v_f,\vp; v_i,0) + K_f
A_{M-1}(\v_f,\vp; v_i,0) =0 \, .\ee
Using the expression of the partial amplitudes $A_M$ we are then led
to ask if
\begin{align}
\sum_{\substack{\v_{M-2},\ldots,\v_{1} \\
\nu_m \neq \nu_{m+1}}} \Big[ \sum_{\substack{\v_{M-1} \\
\nu_{M-1} \neq \nu_{M-2}}}(-i\partial_{\vp} + \wh{D}_f)\,
A_M(\nu_f, \nu_{M-1}, \ldots, \v_1, \v_i; \varphi)
+ \wh{K}_f A_{M-1}(\nu_f, \nu_{M-2}, \ldots, \v_1, \v_i; \varphi) \Big]
\end{align}
vanishes for each $M$. Using the expression (\ref{partialamp}) of
$A(\v_f,v_{M-1}, \ldots \v_1,\v_i; \vp)$, one can readily verify
that this is indeed the case. As in section \ref{s3.3}, the equation
is satisfied `path by path', i.e., already by the intermediate
amplitudes $A(\v_f,v_{M-1}, \ldots \v_1,\v_i; \vp)$ and
$A(\v_f,v_{M-2}, \ldots \v_1,\v_i; \vp)$.

Thus we have shown that the vertex expansion resulting from the
perturbation series satisfies quantum dynamics in a well-controlled
fashion: If we were to terminate the sum at $M=M^\star$, we would
have
\be (i\partial_{\vp} + D_f +\l K)\, \Big[\sum_{M=0}^{{M}^\star}\, \l^M
A_M(\v_f,\vp; v_i,0)\Big] = \mathcal{O}(\l^{{M}^\star+1}) \ee
This brings out the precise sense in which a truncation to a finite
order of the vertex expansion incorporates the quantum dynamics of
the deparameterized theory approximately.

\section{Discussion}
\label{s5}

Because LQC is well-developed in the Hamiltonian framework, it
provides an interesting avenue to probe various aspects of the spin
foam paradigm. For definiteness we focused on the Friedmann model
with a massless scalar field as source. We used the group averaging
procedure that is available for general constrained systems as well
as the natural deparametrization, with $\p$ as the emergent time
variable, that is often employed in LQC.

Group averaging provides a Green's function $G(\v_f,\p_f;
\v_i,\p_i)$ representing the inner product between physical states
extracted from the kinematic kets $|v_f,\p_f\ket$ and
$|\v_i,\p_i\ket$. The Schr\"odinger evolution of the deparameterized
theory provides the transition amplitude $A(\v_f,\p_f; \v_i,\p_i)$
for the physical state $|\v_i\ket$ at the initial instant $\p_i$ to
evolve to the state $|\v_f\ket$ at the final instant of time $\p_f$.
We saw in section \ref{s4} that the two quantities are equal. But
they emphasize different physics. Following the general procedure
invented by Feynman to pass from a Hamiltonian theory to a sum over
histories, we were able to obtain a series expansion for each of
these quantities ---Eq (\ref{final}) for $G(\v_f,\p_f; \v_i,\p_i)$
and Eq (\ref{finalamp}) for $A(\v_f,\p_f; \v_i,\p_i)$--- that mimic
the vertex expansion of SFMs. In section \ref{s3}, we had to make
one assumption in the derivation of the vertex expansion of
$G(\v_f,\p_f; \v_i,\p_i)$: in the passage from (\ref{alambda}) to
(\ref{phy4}) we assumed that the integration over $\alpha$ of the
group averaging procedure commutes with an infinite sum in
(\ref{alambda}). Since the integration over $\alpha$ is by-passed in
the deparameterized framework this assumption was not necessary in
our derivation of the vertex expansion of $A(\v_f,\p_f; \v_i,\p_i)$
in section \ref{s4}.

Detailed parallels between our construction and SFMs are as follows.
The analog of the manifold $M$ with boundaries $S_i,S_f$ in SFMs is
the manifold $\V\times\I$, where $\V$ is the elementary cell in LQC
and $\I$, a closed interval in the real line (corresponding to $\tau
\in [0,1]$ in the timeless framework and $\p \in [\p_f,\p_i]$ in the
deparameterized). The analog of a triangulation in spin-foams is
just a division of $\V\times\I$ into $M$ parts by introducing $M-1$
time slices. Just as the triangulation in SFMs is determined by the
number of 4-simplices, what matters in LQC is the number $M$; the
precise location of slices is irrelevant. The analog of the
dual-triangulation in SFMs is just a `vertical' line in $\V\times\I$
with $M$ marked points or `vertices' (not including the two
end-points of $\I$). Again, what matters is the number $M$; the
precise location of vertices is irrelevant.
%
%\footnote{This is in contrast with the situation in the kinematical
%Hilbert space where the label $\mu$ used in LQC captures both the
%$j$ label carried by an edge as well as its parameter length
%\cite{abl}.}
%
Coloring of the dual-triangulation in SFMs corresponds to an ordered
assignment $(\v_M,\v_{M-1}, \ldots \v_1,\v_0)$ of volumes to edges
bounded by these marked points (subject only to the constraints
$\v_M= \v_f,\,\, \v_0=\v_i$ and $\v_m \not= \v_{m-1}$). Each vertex
signals a change in the physical volume along the quantum history.
\footnote{In the Bianchi models there are additional labels 
corresponding to anisotropies \cite{chn}. These are associated with the faces of the dual graph, and are thus analogs of the spin labels $j$ associated with faces of general spin foams.}
The probability amplitude associated with the given coloring is
given by $A(\v_f, \ldots ,\v_0;\p_f,\p_i)$ in the group averaging
procedure (see Eq (\ref{physamp})) and by $A(\v_f, \ldots
,\v_0;\varphi)$ in the deparametrization procedure (see Eq
(\ref{partialamp})). A sum over colorings yields the partial
amplitude associated with the triangulation with $M$ `vertices'. The
Green's function $G(\v_f,\p_f; \v_i,\p_i)$ and the total transition
amplitude $A(\v_f,\varphi;\,\v_i,0)$ are given by a sum over these
$M$-vertex amplitudes.

Thus, the physical inner product of the timeless framework and the
transition amplitude in the deparameterized framework can each be
expressed as a \emph{discrete sum} without the need of a `continuum
limit': A countable number of vertices suffices; the number of
volume transitions does not have to become continuously infinite.
This result supports the view that LQG and SFMs are not quite
analogous to quantum field theories on classical space-times.
Discrete quantum geometry at the Planck scale makes a key
difference. In sections \ref{s3.2} and \ref{s4.2} we were able to
obtain the same vertex sum using a perturbative expansion, in a
coupling constant $\lambda$, that is reminiscent of GFTs. In
sections \ref{s3.3} and \ref{s4.3} we showed that this is a useful
expansion in the sense that the Green's function and the transition
amplitude satisfy the dynamical equations order by order in
$\lambda$. Thus, if we were to truncate the expansion to order $M$,
the truncated Green's function and transition amplitude would
satisfy the dynamical equations up to terms of the order
$O(\lambda^{M+1})$. Finally in section \ref{s3.4} we showed that the
coupling constant $\lambda$ inspired by GFTs is closely related to
the cosmological constant. This interpretation opens a possibility
that a detailed study of the renormalization group flow in GFT may
be able to account for the very small, positive value of the
cosmological constant.

Taken together, these results provide considerable concrete support
for the general paradigms that underlie SFM and GFT.%
\footnote{But it also brings out the fact that the term `third
quantization' that is sometimes used in GFTs can be misleading
in other contexts. In cosmology, the term is often used to signify 
a Fock space of universes, where the `single universe sector' is
described by the theory presented here.}
However, we emphasize that this analysis has a key limitation: We
did not begin with a SFM in full general relativity and then arrive
at the LQC model through a systematic symmetry reduction of the full
vertex expansion. Rather, we began with an already symmetry reduced
model and recast the results in the spin foam language.
Reciprocally, a key strength of these results is that we did not
have to start by \emph{postulating} that the physical inner product
or the transition amplitude is given by a formal path integral.
Rather, a rigorously developed Hamiltonian theory guaranteed that
these quantities are well-defined. We simply recast their
expressions as vertex expansions. The complementarity of the two 
methods is brought to forefront in the recent work \cite{brv} on
spin-foams in the cosmological context. There, one begins with general spin foams, introduces homogeneity and isotropy only as a restriction on the boundary state and calculates just the leading order terms in the vertex expansion. By contrast, in this work we restricted ourselves to homogeneity and isotropy at the outset but calculated the physical inner product (or, in the deparameterized picture, the transition amplitude) to all orders in the vertex expansion.
 
It is often the case that exactly soluble models not only provide
support for or against general paradigms but they can also uncover
new issues whose significance had not been realized before. The LQC
analysis has brought to forefront three such issues.

First, it has revealed the advantage of adding matter fields. It is
widely appreciated that on physical grounds it is important to
extend SFMs beyond vacuum general relativity. However what was not
realized before is that, rather than complicating the analysis, this
generalization can in fact lead to interesting and significant
technical simplifications. This point is brought out vividly by a
recent analysis of Rovelli and Vidotto \cite{rv}. They considered a
simple model on a finite dimensional Hilbert space where there is no
analog of the scalar field or the possibility of deparametrization.
There, individual terms in the vertex expansion turn out to be well
defined only after a (natural) regularization. In our example, the
presence of the scalar field simplified the analysis (in the
transition from (\ref{phy3}) to (\ref{physamp})) and individual
terms in the vertex expansion are finite without the need of any
regularization. Furthermore, this simplification is not an artefact
of our restriction to the simplest cosmological model. For example,
in the Bianchi I model the Hamiltonian theory is also well-developed
in the \emph{vacuum} case \cite{madrid}. Work in progress
by Campiglia, Henderson, Nelson and Wilson-Ewing shows that
technical problems illustrated in \cite{rv} arise also in
this case, making it necessary to introduce a regularization. These
problems simply disappear if one also includes a scalar field. A qualitative argument suggests that the situation would be similar  beyond cosmological models as well.

Second, it came as a surprise that there are two \emph{distinct}
vertex expansions: Group averaging provides one that mainly uses the
matrix elements of $\Theta$ while the deparameterized framework
provides one that uses only the matrix elements of $\sqrt{\Theta}$.
This is not an artefact of using the simplest cosmological model.
Work in progress indicates that the situation is similar in the
anisotropic Bianchi models. Indeed, from a Hamiltonian perspective,
it would appear that distinct vertex expansions can arise whenever a
well-defined deparametrization is available. This raises an
interesting and more general possibility. Can there exist distinct
spin foam models ---constructed by using, say, distinct vertex
amplitudes---  for which the complete vertex expansions yield the
same answer? Finite truncations of these expansions could be
inequivalent, but each could be tailored to provide an excellent
approximation to the full answer for a specific physical question.
One may then be able to choose which truncated expansion to use to
probe a specific physical effect.

The third issue concerns three related questions in the spin foam
literature: i) Should the physical inner products between states
associated with spin networks be real rather than complex
\cite{ba-ch}? ii) In the classical limit, should one recover $\cos
S$ in place of the usual term $e^{iS}$, where $S$ is the Einstein
Hilbert action \cite{limit1,limit2}? iii) Should the choice of
orientation play a role in the sum of histories \cite{do}? In the
LQC example we studied in this paper, these three questions are
intimately related. The inner product between the physical states
$[\v,\p]_+$ determined by the kinematic basis vectors ---which are
the analogs of spin networks in this example--- are in general
complex (see Eq (\ref{final})). However, if we had dropped the
positive frequency requirement, the group averaged inner products
would have been real (see Eq (\ref{phy3})). The situation with
action is analogous. And, as we show in the next paragraph, the
positive frequency condition also selects a time-orientation.

Since this is an important issue, we will discuss it in some detail.
Let us begin with the classical theory. The phase space is
4-dimensional and there is a single constraint: $C (\v,b;\p ,p_\p)
:= G\,p_\p^2 - 3\pi\,(\lp^2 \v^2)\, b^2 =0$. Dynamics has two
conceptually interesting features. First, given a solution $(\v(t),
\p(t))$ to the constraint and dynamical equations, $(-\v(t), \p(t))$
is also a solution (where $t$ denotes proper time). They define the
same space-time metric and scalar field; only the parity of the
spatial triad is reversed. Therefore $(\v(t), \p(t)) \rightarrow
(-\v(t), \p(t))$ is regarded as a gauge transformation. The second
feature arises from the fact that the constraint surface has two
`branches', $p_\p >0$ and $p_\p <0$, joined at points $p_\p =0$
which represent Minkowski space-time. As is usual in quantum
cosmology, let us ignore the trivial flat solution. Then each of the
two portions $\bar\Gamma^\pm$ of the constraint surface defined by
the sign of $p_\p$ is left invariant by dynamics. Furthermore, there
is a symmetry: Given a dynamical trajectory $(\v(t), \p(t))$ in
$\bar\Gamma^+$, there is a trajectory $(\v(t), -\p(t))$ which lies
in $\bar\Gamma^-$. This represents a \emph{redundancy} in the
description in the sense that we recover all physical space-time
geometries $g_{ab}(t)$ even if we restrict only to one of the two
branches $\bar\Gamma^\pm$. In particular, the dynamical trajectories
on $\bar\Gamma^+$, for example, include solutions which start with a
big-bang and expand out to infinity \emph{as well as} those which
start out with infinite volume and end their lives in a big crunch.
The difference is in only in time orientation: If we regard $\phi$
as an internal or relational time variable and \emph{reconstruct}
space-time geometries from phase space trajectories, space-times
obtained from a trajectory on $\bar\Gamma^+$ defines the same
geometry as the one obtained  from the corresponding trajectory on
$\bar\Gamma^-$ but with opposite time orientation. As in the
Klein-Gordon theory of a free relativistic particle, this redundancy
is removed by restricting oneself either to the $p_\p >0$ sector or
to the $p_\p <0$ sector. In the quantum theory, then, the physical
Hilbert space is given by solutions $\Psi(\v,\p)$ to the quantum
constraint (\ref{qc}) which in addition have only positive (or
negative) frequency so that the operator $p_\p$ is positive (or
negative) definite. (They are also invariant under parity,
$\Psi(\v,\p) = \Psi(-\v, \p)$). Thus, the LQC example suggests that
in general SFMs one should fix the time-orientation, lending
independent support to the new ideas proposed in \cite{do}. Reality
of the physical inner products between spin network states
\cite{ba-ch} and the emergence of $\cos S$ in place of $e^{iS}$
\cite{limit1,limit2} can be traced back to the fact that in most of
the SFM literature one sums over both orientations. However, our
analysis provides only a hint rather than an iron-clad argument
because all our discussion is tied to LQC models where symmetry
reduction occurs prior to quantization.

We conclude with an observation. We have recast LQC as a sum over
histories. However, this is different from a Feynman path integral
in which the integrand is expressed as $e^{iS}$, for a suitable
action $S$. This step was not necessary for the goals of this paper.
However, it is of considerable interest, especially in the
cosmological context, for certain physical issues such as the
emergence of the classical universe and semi-classical corrections
to the classical theory. Such a path integral formulation of LQC
does exist \cite{ah} and will be discussed elsewhere.%
\footnote{A path integral formulation of polymer quantum mechanics
was carried out independently by Husain and Winkler \cite{hw}.}

\section*{Acknowledgments} We would like to thank Jerzy Lewandowski,
Daniele Oriti, Vincent Rivasseau and Carlo Rovelli for discussions
and Laurent Freidel and Kirill Krasnov for their comments. This work
was supported in part by the NSF grant PHY0854743 and the Eberly
research funds of Penn State.

\appendix

\section{Limit in Eq (\ref{lim1})}
\label{prooflimit}

It is convenient to rewrite $A_N(\v_M, \ldots, \v_0; \alpha)$
defined in (\ref{AN1}) in the following way:
\begin{align}
 A_N(\v_M, \ldots, \v_0; \alpha) &= U_{\nu_M \nu_{M-1}}
\ldots U_{\nu_1 \nu_0}\left[U_{\nu_{M} \nu_{M}}\right]^N
\left[U_{\nu_M \nu_{M}}\ldots U_{\nu_0 \nu_{0}}\right]^{-1}
\,\, \times \nonumber \\
& \sum_{N_M=M}^{N-1} \; \sum_{N_{M-1}=M-1}^{N_M-1}
\ldots \sum_{N_1=1}^{N_2-1} \;  \left[\frac{U_{\nu_{M-1}
\nu_{M-1}}}{U_{\nu_{M}\nu_{M}}}\right]^{N_M} \ldots
\left[\frac{U_{\nu_{0} \nu_{0}}}{U_{\nu_{1} \nu_{1}}}\right]^{N_1}.
\label{AN2}
\end{align}
Our aim is to calculate the limit $N \rightarrow \infty$ of
(\ref{AN2}) and show that is given by  $A(\v_M, \ldots, \v_0;
\alpha)$, of Eq (\ref{lim1}) which we rewrite as
\begin{align}
 A(\v_M, \ldots, \v_0; \alpha) &= (-i \alpha )^M\,
\Theta_{\v_M \v_{M-1}} \ldots  \Theta_{\v_1 \v_0}\, e^{-i \alpha
\Theta_{\v_M \v_M}}\,\, \times \nonumber \\
& \qquad \sint_0^1 \dd\tau_M\, \sint_0^{\tau_M} \dd\tau_{M-1}\,
\ldots \sint_0^{\tau_2} \dd\tau_{1}\,\,\, e^{\tau_M b_M}\ldots
e^{\tau_1 b_1}  \label{A} \end{align}
where \be \label{bm} b_m:=-i \alpha (\Theta_{\nu_{m-1} \nu_{m-1}} -
\Theta_{\nu_{m} \nu_m}). \ee

We start by calculating the $N\gg 1$ behavior of the terms appearing
in (\ref{AN2}). These are:
\ba U_{\nu_{m+1}\nu_{m}}  &  = &  -\frac{i \alpha}{N}
\Theta_{\nu_{m+1} \nu_m} + O(N^{-2}), \label{term1}\ea
\ba \left[U_{\nu_{M} \nu_{M}}\right]^N  & = & e^{N \log U_{\nu_{M}
\nu_{M}}} \nonumber \\
& = &   e^{N\left(-i \frac{\alpha}{N} \Theta_{\nu_{M} \nu_{M}}+
O(N^{-2})\right)} \nonumber \\
& = & e^{-i \alpha \Theta_{\nu_{M} \nu_{M}}} + O(N^{-1}),
\label{term2} \ea
\ba \left[U_{\nu_M \nu_{M}}\ldots U_{\nu_0 \nu_{0}}\right]^{-1} & =&
1 + O(N^{-1}), \label{term3} \ea \ba \left[\frac{U_{\nu_{m-1}
\nu_{m-1}}}{U_{\nu_{m} \nu_{m}}}\right]^{N_m} & = & e^{N_m
\left(\log U_{\nu_{m-1} \nu_{m-1}}- \log U_{\nu_{m} \nu_{m}}
\right)} \nonumber \\
& = & e^{N_m \left(b_m/N +O(N^{-2)} \right)} \nonumber \\
& = & e^{\frac{N_m}{N} b_m}+O(N_m N^{-2}) , \label{term4} \ea
with $b_m$ given in (\ref{bm}). In (\ref{term2}) and (\ref{term4})
we have used the fact that the multivalued nature of the $\log$
function does not affect the final result: $e^{N(\log x + 2 \pi i
k)}=e^{N \log x}$ where $k \in \Z$ reflects the multiple values that
$\log$ can take.

We now  substitute expressions (\ref{term1}) to (\ref{term4}) in
(\ref{AN2}) to obtain
\begin{align}
A_N(\v_M, \ldots, \v_0; \alpha) =& \left[(-i \alpha )^M
\Theta_{\v_M \v_{M-1}} \ldots \Theta_{\v_1 \v_0}
e^{-i \alpha \Theta_{\v_M \v_M}} N^{-M}+O(N^{-M-1}) \right]\,
\times \nonumber\\
&\prod_{m=1}^M \left[ \sum_{N_m=m}^{N_{m+1}-1} e^{\frac{N_m}{N} b_m}+
O(N_m N^{-2}) \right]  \label{A2}
\end{align}
where the product denotes the $M$ nested sums in (\ref{AN2}). Each
sum in (\ref{A2}) has two terms. The first one gives a contribution
of $\sum_{N_m} e^{\frac{N_m}{N} b_m} \sim O(N)$ while the second one
is $\sum_{N_m} O(N_m N^{-2}) \sim O(1)$. The $M$ sums then give a
contribution of order $[O(N)+O(1)]^M \sim O(N^M)+O(N^{M-1})$.  By
combining this with the first factor of (\ref{A2}), we find that the
non-vanishing contribution comes from the first terms of the sums:
\begin{align}
A_N(\v_M, \ldots, \v_0; \alpha) =& (-i \alpha )^M\,
\Theta_{\v_M \v_{M-1}}\, \ldots\, \Theta_{\v_1 \v_0}\,
e^{-i \alpha \Theta_{\v_M \v_M}}\, \times \nonumber\\
& N^{-M}\,\prod_{m=1}^M\, \left[\sum_{N_m=m}^{N_{m+1}-1}\,
e^{\frac{N_m}{N}\, b_m} \right] + O(N^{-1}). \label{A3}
\end{align}

Eq (\ref{A3}) has all the pre-factors appearing in (\ref{A}). It
then remains to show that  $N^{-M}$ times the sums in (\ref{A3})
limits to the integrals in (\ref{A}). But this is rather obvious, as
the sums can be seen as Riemann sums for the integrals.
Specifically,
\begin{align}
& \lim_{N \to \infty}N^{-M} \prod_{m=1}^M \left[ \sum_{N_m=m}^{N_{m+1}-1}
e^{\frac{N_m}{N} b_m} \right]  &    \nonumber \\
& = \lim_{N \to \infty} N^{-M}\sum_{N_M=0}^{N} \;
\sum_{N_{M-1}=0}^{N_M} \ldots \sum_{N_1=0}^{N_2}
e^{\frac{N_M}{N} b_M} \ldots e^{\frac{N_1}{N} b_1} & \nonumber \\
& = \sint_0^1 \dd\tau_M\, \sint_0^{\tau_M} \dd\tau_{M-1}\,
\ldots \int_0^{\tau_2} \!\dd\tau_{1}\,\,\, e^{\tau_M b_M}
\,\ldots\, e^{\tau_1 b_1}\, &
\end{align}
where, in the second line, we have slightly changed the limits on
the sums, introducing an $O(N^{-1})$-term which vanishes in the
limit. This concludes the proof of the limit (\ref{lim1}).

\section{General Integrals in Eq (\ref{lim1})}
\label{gralintegral}

The integrals over $\tau$ appearing in the amplitude for a single
discrete path (3.18) can be evaluated for a general sequence of
volumes $(\v_M,...,\v_0)$ with the result given by (3.21).  In this
appendix we will perform these integrals first for the case where
all $\v_i$ are distinct and then for the general case.  The
amplitude for a single discrete path given by (3.18) and (3.19) is
\begin{align}
A(\nu_M, \ldots, \nu_0, \alpha) =&  \sint_0^{\Delta \tau}
\dd\tau_M\, \sint_0^{\tau_M} \dd\tau_{M-1}\, \ldots \sint_0^{\tau_2}
\dd\tau_{1} e^{-i(\Delta \tau - \tau_M) \alpha \Theta_{\v_M\v_M}}\,\,
 (-i \alpha \Theta_{\v_M\v_{M_1}})\,\, \times \nonumber\\
& e^{-i(\tau_M-\tau_{M-1})\alpha \Theta_{\v_{M-1}\v_{M-1}}}\,
\ldots\,\, e^{-i(\tau_2-\tau_1)\alpha \Theta_{\v_1\v_1}}\,\,
(-i \alpha \Theta_{\v_1\v_{0}})\,\, e^{i\tau_1 \alpha \Theta_{\v_0\v_0}}
\end{align}
This expression can be written in terms of the following integral.
\begin{align}
I(x_M,\ldots,x_0, \Delta \tau) = \sint_0^{\Delta \tau}
\dd\tau_M\, \sint_0^{\tau_M} \dd\tau_{M-1}\, \ldots
\sint_0^{\tau_2} \dd\tau_{1} (i)^M
\,e^{i (\Delta \tau- \tau_M) x_M}\,
e^{i (\tau_M - \tau_{M-1}) x_{M-1}} \\
\nonumber... e^{i (\tau_2-\tau_1) x_1} e^{i \tau_1 x_0}
\end{align}

We will first evaluate this integral for the case where all $x_i$
are distinct.  By induction on $M$ ---the number of vertices or the
number of times that $x$ changes value--- we will show that when the
$x_i$ are all distinct the integral is given by
 \begin{align}
\label{distinct} I(x_M, \ldots ,x_0, \Delta \tau) = \sum_{i=0}^M
\frac{e^{i x_i \Delta \tau}}{\prod_{j \neq i}^M (x_i - x_j)}
\end{align}
This is true by inspection for $M=0$.  If we assume that
(\ref{distinct}) holds for $M$ we can evaluate the integral with
$M+1$ vertices.
\begin{align}
I(x_{M+1},x_M,\ldots,x_0, \Delta \tau) &= \sint_0^{\Delta \tau}
\dd\tau_{M+1} \;ie^{i (\Delta \tau- \tau_{M+1}) x_{M+1}}
I(x_M,\ldots, x_0, \tau_{M+1}) \\
\nonumber & =\sint_0^{\Delta \tau} \dd\tau_{M+1}\,
i e^{i (\Delta \tau- \tau_{M+1}) x_{M+1}} \sum_{i=0}^M
\frac{e^{i x_i  \tau_{M+1}}}{\prod_{j \neq i}^M (x_i - x_j)} \\
\nonumber & = \sum_{i=0}^M
\frac{e^{i x_i  \Delta \tau}}{\prod_{j \neq i}^{M+1} (x_i - x_j)}
- e^{i \Delta \tau x_{M+1}} \sum_{i=0}^M \frac{1}{\prod_{j \neq i}^{M+1}
 (x_i - x_j)}
\end{align}
In the first step we recognized that the $M+1$-th integral contains
the $M$-th and then, in the second step, we inserted the assumed
result for the $M-th$ integral.  In the second step the integral
over $\tau_{M+1}$ is carried out.  Finally using the identity
\be \label{identa} \sum_{i=1}^{M+1} \frac{1}{\prod_{j\neq i}^{M+1}
(x_i - x_j)}=0 \ee
The integral can be written as
\begin{align}
I(x_{M+1},x_M,\ldots,x_0, \Delta \tau) &=  \sum_{i=0}^{M+1}
\frac{e^{i x_i  \Delta \tau}}{\prod_{j \neq i}^{M+1} (x_i - x_j)}
\end{align}
Therefore if (\ref{distinct}) holds for $M$ it also holds for $M+1$,
thus by induction it holds for all $M \geq 0$.

If the $x_i$ are not distinct, if there exist $i,j$ such that $x_i =
x_j$, then the proof follows in a similar way.  The key element is
that the integral $I(x_M,...,x_0)$ is independent of the order of
the $x_i$'s.  This can be seen by rewriting the integral in terms of
the time intervals $\Delta \tau_i = \tau_{i+1}-\tau_i$ where
$\tau_0=0$ and $\tau_{m+1} = \Delta \tau$. %%
\begin{align}
I(x_0,x_1,...x_M, \Delta \tau) = \sint_0^{\Delta \tau} \dd \Delta
\tau_M \dd \Delta \tau_{M-1}.. \dd \Delta \tau_{0}\,\, \delta
(\Delta \tau_{m} + ... + \Delta \tau_{0}-\Delta \tau) \\
\nonumber (i)^M e^{i \Delta \tau_{M} x_M} e^{i \Delta \tau_{M-1}
x_{M-1}} ... e^{i \Delta \tau_1 x_1} e^{i \Delta \tau_0 x_0}
\end{align}
It is clear that this is symmetric under the interchange of $x_i$
with $x_j$ for all $i,j$, so the integral is independent of the
order of the sequence $x_i$.  Since the integral is independent of
the order of the values $x_i$ it should be characterized by the
distinct values, labeled by $y_i$ and their multiplicity $n_i$.
Where $n_1+ \ldots + n_p=M+1$.  Given a set of values $x_i$ we will
evaluate the integral for the case where they are organized such
that any $x_i$ sharing the same value are grouped together. Doing so
the integral simplifies to
\begin{align}
\label{general} I(y_p,n_p, \ldots ,y_1,n_1, \Delta \tau) =
\sint_0^{\Delta \tau} \dd\tau_M\, \sint_0^{\tau_M} \dd\tau_{M-1}\,
\ldots \sint_0^{\tau_2} \dd\tau_{1} (i)^M e^{i (\Delta \tau -
\tau_{n_1+...+n_{p-1}}) y_p} \\
\nonumber e^{i (\tau_{n_1+...+n_{p-1}} - \tau_{n_1+...+n_{p-2}})
y_{p-1}}... e^{i (\tau_{n_1+n_2}-\tau_{n_1}) y_2} e^{i \tau_{n_1} y_1}
\end{align}

By induction on $p$, the number of distinct values, we show that
this integral is given by
\begin{align}
\label{genres} I(y_p,n_p,...,y_1,n_1, \Delta \tau) &=
\frac{1}{(n_p-1)!}\left(\frac{\partial}{\partial y_p}\right)^{n_p-1}
\ldots \frac{1}{(n_1-1)!}\left( \frac{\partial}{\partial y_1} \right)^{n_1-1}
 \sum_{i=1}^p \frac{e^{i y_i \Delta \tau}}{\prod_{j \neq i}^p (y_i - y_j)} \\
\nonumber &= \prod_{k=1}^p \frac{1}{(n_k-1)!} \left(
\frac{\partial}{\partial y_k} \right)^{n_k-1}
\sum_{i=1}^p \frac{e^{i y_i \Delta \tau}}{\prod_{j \neq i}^p
(y_i - y_j)}
\end{align}
For $p=1$ (\ref{general}) can be easily evaluated giving
\begin{align}
I(y_1,n_1) &= \sint_0^{\Delta\tau} d\tau_{n_1-1} \ldots
\sint_0^{\tau_2} d\tau_1 (i)^{n_1-1} e^{i y_1 \Delta \tau}
= \frac{(i \Delta \tau)^{n_1-1}}{(n_1-1)!}e^{i y_1 \Delta \tau} \\
\nonumber &=\left(\frac{\partial}{\partial y_1}\right)^{n_1-1}
\frac{1}{(n_1-1)!} e^{i y_i \Delta \tau}
\end{align}
If we assume that (\ref{genres}) holds for $p$ distinct values then
we can evaluate it for $p+1$ distinct values as follows.
\begin{align}
I(y_{p+1},n_{p+1}, y_p, n_p \ldots ,y_1,n_1, \Delta \tau)
= \sint_0^{\Delta \tau} \dd\tau_{M}\, \ldots
\sint_0^{\tau_{M-n_{p+1}+2}} \dd\tau_{M-n_{p+1}+1} \\
\nonumber  (i)^{n_{p+1}-1}  e^{i (\Delta \tau -
\tau_{M-n_{p+1}+1}) y_{p+1}}  I(y_p,n_p, \ldots, y_1, n_1,
\tau_{M-n_{p+1}+1})
\end{align}
Plugging in the assumed result for $p$ distinct values and
performing the integrals over $\tau$ we obtain
\begin{align}
I(y_{p+1},n_{p+1}, \ldots ,y_1,n_1, \Delta \tau) &=
\prod_{k=1}^p \frac{1}{(n_k-1)!} \left( \frac{\partial}{\partial y_k}
\right)^{n_k-1} \sum_{i=1}^p \frac{1}{\prod_{j \neq i}^p (y_i - y_j)}\\
\nonumber &\left[ \frac{e^{i y_i \Delta\tau}}{(y_i-y_{p+1})^{n_{p+1}}}
- \sum_{m=0}^{n_{p+1}} \frac{e^{i y_{p+1} \Delta \tau}}{(y_i-y_{p+1})^m}
\frac{(i \Delta \tau)^{n_{p+1}-m}}{(n_{p+1}-m)!}\right]
\end{align}
We recognize that the term in brackets can be written as derivatives
with respect to $y_{p+1}$ of a simple function.
\begin{align}
I(y_{p+1},n_{p+1}, y_p, n_p \ldots ,y_1,n_1, \Delta \tau)=
\prod_{k=1}^p \frac{1}{(n_k-1)!} \left( \frac{\partial}{\partial y_k}
\right)^{n_k-1} \sum_{i=1}^p \frac{1}{\prod_{j \neq i}^p (y_i - y_j)} \\
\nonumber \left[\frac{1}{(n_{p+1}-1)!} \left(
\frac{\partial}{\partial y_{p+1}} \right)^{n_{p+1}-1}
\left( \frac{e^{i y_i \Delta \tau}}{y_i - y_{p+1}} -
\frac{e^{i y_{p+1} \Delta \tau}}{y_i - y_{p+1}} \right) \right]
  \end{align}
Finally simplifying the expression and using eqn (\ref{identa}) we
obtain
  \begin{align}
I(y_{p+1},n_{p+1},  \ldots ,y_1,n_1, \Delta \tau)=
\prod_{k=1}^{p+1} \frac{1}{(n_k-1)!} \left(
\frac{\partial}{\partial y_k} \right)^{n_k-1}
\sum_{i=1}^{p+1} \frac{e^{i y_i \Delta \tau}}{\prod_{j \neq i}^p
(y_i - y_j)}
\end{align}
Thus if (\ref{genres}) holds for $p$ then it also holds for $p+1$,
so it is true for all $p \geq 0$.  Using this result we find that
the contribution due to each discrete path is
\begin{align}
A(\nu_M, \ldots, \nu_0, \alpha) = (\Theta_{\v_M\v_{M-1}})
(\Theta_{\v_{M-1}\v_{M-2}}) \ldots ( \Theta_{\v_2\v_{1}})
( \Theta_{\v_1\v_{0}})\\
\nonumber \prod_{k=1}^p \frac{1}{(n_k-1)!} \left(
\frac{\partial}{\partial  \Theta_{w_k w_k}} \right)^{n_k-1}
\sum_{i=1}^p \frac{e^{-i \alpha \Theta_{w_iw_i}\Delta \tau}}
{\prod_{j \neq i}^p ( \Theta_{w_iw_i} -  \Theta_{w_j w_j})}
\end{align}
where $w_i$ label the distinct values taken by $\v$ along the path
and $n_i$ the multiplicity of each value.

\section{Eigenstates and Operator functions of ${\Theta}$}
\label{matrixel}

In the timeless framework of section \ref{s3}, the vertex expansion
mostly featured matrix elements $\Theta_{\v_m\v_n} = \bra \v_m|
\Theta | \v_n \ket$. These are easy to evaluate directly from the
definition (\ref{theta}) of $\Theta$. In the deparameterized
framework of section \ref{s4}, on the other hand, the vertex
expansion involves matrix elements of $\sqrt{\Theta}$. To evaluate
these one needs the spectral decomposition of $\Theta$. In the first
part of this Appendix we construct eigenstates of $\Theta$ and
discuss their relevant properties. In the second part we use these
eigenstates to evaluate the matrix elements functions of $\Theta$,
including $\sqrt{\Theta}$.

\subsection{Eigenstates of $\Theta$}

Recall that $\Theta$ is a positive, self-adjoint operator on $\Hkg$.
By its definition (\ref{theta}), it follows that $\Theta$ preserves
each of the three sub-spaces in the decomposition $\Hkg = \H_-
\oplus \H_0 \oplus \H_+$, spanned by functions with support on
$\nu<0$, $\nu=0$ and $\nu >0$ respectively. In particular, $|\nu=0
\ket$ is the unique eigenvector of $\Theta$, with eigenvalue $0$;
$\H_0$ is 1-dimensional. Our first task is to solve the eigenvalue
equation for a general eigenvalue $\w_k^2$:
\be \label{evalueeq} \Theta\, e_k(\nu)= \w_k^2  \, e_k(\nu)\,. \ee
This task becomes simpler in the representation in which states are
functions $\chi(b)$ of the variable $b$ conjugate to $\v$:
\footnote{Our normalization is different from that in \cite{acs}.
The wave function $\tilde{\Psi}(\nu)$ in \cite{acs} is related to
the one here by $\Psi(\nu)= \sqrt{\frac{\lo}{\pi |\nu|}}\,
\tilde{\Psi}(\nu)$.}
\be \label{fourier} \chi(b) := \sqrt{\frac{\lo}{\pi}}\,
\sum_{\nu=4n\lo} e^{\frac{i}{2}\,\v b} \,\,
\f{\Psi(\v)}{\sqrt{|\nu|}} \, . \ee
In this representation, the eigenvalue equation (\ref{evalueeq})
takes the form of a simple differential equation
\be \big( \Theta \chi_k \big) (b) = -12 \pi G \left(\frac{\sin\lo
b}{\lo} \partial_b\right)^2 \chi_k(b) = \w_k^2 \, \chi_k(b), \ee
whose solutions are
\be \label{estateb} \chi_k (b) = A(k)\, e^{ik \log(\tan \frac{\lo
b}{2})} \qquad {\rm with}\quad \w_k^2= 12 \pi G \, k^2\, , \ee
where $A(k)$ is a normalization factor and $k \in (-\infty,
\infty)$. $k=0$ yields a discrete eigenvalue $\w_k=0$ and in the
$\v$ representation the eigenvector can be expressed simply as
$e_0(\nu)=\delta_{0, \nu}$. Eigenvectors with non-zero eigenvalues
can also be expressed in the $\v$ representation by applying the
inverse transformation of (\ref{fourier}) to (\ref{estateb}):
\be \label{psik} e_k(\v)  = A(k) \sqrt{\frac{\lo |\v|}{\pi}}
\sint_0^{\pi/\lo} \dd b\; e^{-\frac{i}{2} \v b}\, e^{i k \log(\tan
\frac{\lo b}{2})} \qquad {\rm where} \quad k \neq 0\, . \ee

Let us note two properties of these eigenvectors. First, $e_{k}$ and
$e_{-k}$ have the same eigenvalue and so the $\w_k^2$-eigenspace is
two-dimensional. Second, the vectors $e_k(\v) $  we have obtained
have support on both $\v>0$ and $\v <0$. However, since $\Theta$
preserves the sub-spaces $\H_\pm$, it is natural to seek linear
combinations $e^\pm_k(\v)$ of $e_k(\v)$ and $e_{-k}(\v)$ which lie
in these sub-spaces. In particular, this will simplify the problem
of normalization of eigenfunctions.

Let us begin by rewriting the integral in (\ref{psik}) as a contour
integral in the complex plane. Recalling that $\v=4\lo\,n$ and
setting $z=e^{i b \lo}$ we obtain
\be \frac{\lo}{\pi} \sint_0^{\pi/\lo} \dd b\; e^{-2 i b n} e^{i k
\log(\tan \frac{\lo b}{2})}= \frac{e^{-\pi k/2}}{\pi i}
\sint_{\mathcal{C}} z^{-2n-1}\left(\frac{1-z}{1+z} \right)^{i k} \dd
z =: J(k,n), \ee
where $\mathcal{C}$ is the unit semicircle in counterclockwise
direction in the upper half, $\Im{z}>0$, of the complex plane. As
remarked earlier, $e_k(\v)= A(k) \sqrt{\lo|\v|/\pi}\,J(k,\v/4\lo)$
has support on both positive and negative values of $\v=4\lo\,n$.
Now, the second independent eigenfunction $e_{-k}(\v)$ with the same
eigenvalue $\w_k^2$ can be represented in a similar fashion by
setting $z=-e^{i b \lo}$. The result is a contour-integral along the
unit semicircle in counterclockwise direction in the lower half,
$\Im{z}<0$ of the complex plane. By combining the two integrals, we
obtain a closed integral along the unit circle:
\be \frac{1}{2\pi i} \oint z^{-2n-1}\left(\frac{1-z}{1+z} \right)^{i
k} \dd z = \frac{1}{2} \left( e^{\pi k/2} J(k,n)+e^{-\pi k/2}
J(-k,n) \right) =: I(k,n)\, . \ee
Being a linear combination of $e_k(\v)$ and $e_{-k}(\v)$, this
$I(k,n)$ gives also an eigenfunction of $\Theta$ with eigenvalue
$\w_k^2$. Moreover, using elementary complex analysis, one finds
that \emph{it has support only on positive} $n$:
\be \label{iplus} I(k,n)= \left\{ \begin{array}{ll} \frac{1}{(2n)!}
\left. \frac{d^{2n}}{d s^{2n}} \right|_{s=0}
\left(\frac{1-s}{1+s} \right)^{i k} & \quad n \geq 0 \\
0 & \quad n <0 .\end{array} \right. \ee
Repeating the argument but taking $z=e^{-i b \lo}$ and $z=-e^{-i b
\lo}$  one obtains
\be \frac{1}{2} \left( e^{-\pi k/2} J(k,n)+e^{\pi k/2} J(-k,n)
\right) =  \frac{1}{2\pi i} \oint z^{2n-1}\left(\frac{1-z}{1+z}
\right)^{i k} \dd z=I(k,-n) \ee
which has support only on negative $n$. Thus, the basis we are
looking for is given by
\be \label{knu} e^{\pm}_k(\v) :=  \frac{1}{2} \left( e^{\pm \pi k/2}
e_k(\v) + e^{\mp \pi k/2} e_{-k}(\v) \right) = A(k)\,\,
\sqrt{\frac{\pi |\v|}{\lo}}\,\, I(k,\pm \frac{\v}{4 \lo})\, . \ee
By construction, $e^\pm_k \in \H_\pm$.

Next, let us calculate the normalization of these vectors. It is
convenient to introduce kets $|k \pm \ket$ such that $\bra \v | k
\pm \ket = e^{\pm}_k(\v)$. Then, it is clear that $\bra k'\pm | k
\mp \ket =0$. To calculate the nontrivial inner product, $\bra k'
\pm | k \pm \ket$, let us return to the $b$ representation. There,
the functions describing the states $|k \pm \ket$ are
\be \chi^{\pm}_k(b)= \frac{A(k)}{2} \left( e^{\pm \pi k/2} e^{ik
\log(\tan \frac{\lo b}{2})}+e^{\mp \pi k/2} e^{-ik \log(\tan
\frac{\lo b}{2})} \right) \ee
and their inner product is given by \cite{acs}
\be \label{ipb} \bra k' \pm | k \pm \ket = \sint_0^{\pi/\lo} \dd b
\; |A(k)|^2\,\,\overline{\chi^\pm}_{k'}(b)\, | 2i \partial_b |\,
\chi^\pm_k(b) \ee
where $| 2i \partial_b |$ is the absolute value of the volume
operator $\hat{\v}=2i \partial_b$. Simplification occurs because
$e^\pm_k(\v)$ have support only on positive/negative $\v$ values.
Because of this property, one can replace $|\partial_b|$ in
(\ref{ipb}) by $\pm\partial_b$. The calculation now reduces to a
straightforward integration. The result is
\be \label{normk} \bra k' \pm | k \pm \ket =  |A(k)|^2\, 2 \pi k\,
\sinh(\pi k)\,\, \delta(k',k). \ee
%

%Using this with (\ref{knu}) we can find the normalization of the
%eigenvectors $e_k(\nu)$, which turn out to be

%\be \bra k' | k  \ket = |A(k)|^2 \,4 \pi k \coth(\pi k)\,
%\delta(k',k). \ee

%Thus, by choosing  $A(k)=  \sqrt{\tanh(\pi k)/(4 \pi k)}$, we get
%the standard continuum normalization for the vectors $e_k(\nu)$.

\subsection{Matrix Elements for $f(\Theta)$}

We will now use the eigenbasis $|\pm k\ket$ of $\Theta$ to calculate
the matrix elements $\bra 4 n \lo |f({\Theta})| 4 m \lo \ket$, of
the operators of the form $f(\Theta)$, for a measurable function
$f$. Throughout this section, the normalization factor $A(k)$ is
chosen to be unity. From the normalization condition (\ref{normk})
with $A(k)=1$, we have the following decomposition of the identity:
\be \mathbf{I}=\sint_{0}^{\infty}\frac{\dd k}{2 \pi k \sinh(\pi
k)}\,\,\left(|k +\ket \bra k +|\,+\,|k -\ket \bra k -|\right). \ee
which can be inserted in $\bra 4 n \lo | f(\widehat{\Theta}) | 4 m
\lo \ket$. If $m$ and $n$ have different signs, the result is zero.
It suffices to consider the case where both are positive. By writing
$\bra 4 n \lo |k+\ket$ in terms of derivatives (see equations
(\ref{knu}) and (\ref{iplus})), one obtains
\be \label{matelf} \bra 4 n \lo  | f(\widehat{\Theta})|4 m \lo 
\ket= \frac{2 \sqrt{m n}}{(2n)!(2 m)!} \left. \frac{d^{2 m}}{d s^{2
m}} \frac{d^{2 n}}{d t^{2 n}} \right|_{s=t=0}
F_{f(\Theta)}\left(\frac{1+s}{1-s}\frac{1-t}{1+t} \right) \ee
with $F_{f(\Theta)}$ the `generating function' given by%
\footnote{For a general $f$, integral as defined may diverge.
However the divergent terms (e.g., those which are $x$-independent)
do not contribute to the expression of the matrix element and can
therefore be discarded. This `finite part extraction' is implicit in
going from (\ref{defF}) to (\ref{genfn1}), (\ref{genfn2}) and
(\ref{genfn3}) .}
\be \label{defF} F_{f(\Theta)}(x)= \sint_{0}^\infty \dd k \;
\frac{f(12 \pi G k^2) x^{i k}}{k \sinh(\pi k)}  . \ee
We now give the generating function for $\sqrt{\Theta}$. It is also
useful (at least to check normalization factors) to write down the
generating functions for operators whose matrix elements are known,
namely $\Theta$ and the identity\, ${I}$. These generating functions
are given by,
\ba F_{I}(x) &=& -2 \left(\log(1+x)+\log \Gamma(1/2+i \frac{\log
x}{2 \pi})
\right) \label{genfn1}\\
F_{\sqrt{\Theta}}(x) & = & \sqrt{12 \pi G}\left(\frac{2 i x}{1+x}-
\frac{1}{\pi} \psi(1/2+i \frac{\log x}{2 \pi}) \right) \label{genfn2}\\
F_{\Theta}(x) & = & 12 \pi G \left( \frac{2 x}{(1+x)^2}-
\frac{1}{2\pi^2} \psi'(1/2+i \frac{\log x}{2 \pi}) \right)
\label{genfn3} \ea
where $\Gamma(z)$ is the gamma function, and
$\psi(z)=\Gamma'(z)/\Gamma(z)$ the polygamma function.

In obtaining these functions, it is useful to observe the following
relations among them:
\ba F_{\sqrt{\Theta}}(x) & = & -i \sqrt{12 \pi G} \, x \frac{d}{d x}
F_{I}(x) \\
F_{\Theta}(x) & = & -i \sqrt{12 \pi G} \, x \frac{d}{d x}
F_{\sqrt{\Theta}}(x), \ea
which can be derived from (\ref{defF}).

We will conclude by noting that the matrix elements for the
evolution operator $U(\varphi)=e^{i \varphi \sqrt{\Theta}}$ are easy
to find: From (\ref{defF}) one sees that $F_{U(\varphi)}(x) =
F_{I}(e^{\sqrt{12 \pi G} \varphi} x)$.

\end{document}